\documentclass{pasj00}

\usepackage{graphicx}
\usepackage{epsfig}
\usepackage{graphics}
\usepackage{lscape}

\begin{document}

\SetRunningHead{K. Vierdayanti, S. Mineshige, and Y. Ueda}{GRS 1915+105 at Near-Eddington Luminosity}
\Received{2009/11/02}
\Accepted{2010/00/00}

\title{Probing the Peculiar Behavior of GRS 1915+105 at Near-Eddington Luminosity}

 \author{%
   Kiki \textsc{Vierdayanti}\altaffilmark{1},
   Shin \textsc{Mineshige}\altaffilmark{1},
   and
   Yoshihiro \textsc{Ueda}\altaffilmark{1}}
 \altaffiltext{1}{Department of Astronomy, Kyoto University, Sakyo-ku, Kyoto 606-8502}
 \email{kiki@kusastro.kyoto-u.ac.jp}


\maketitle

\begin{abstract}
To understand the nature of supercritical accretion, we systematically analyze
the {\it RXTE}/PCA data of GRS 1915+105 in its quasi-steady states, by choosing data
with small variability during 1999 -- 2000.
We apply a multicolor disk plus a thermal Comptonization model and take into
consideration accurate interstellar absorption, a reflection component (with
an iron-K emission line), and absorption features from the disk wind
self-consistently. The total luminosity ranges from $\sim 0.2 L_{\rm E}$ to
slightly above $L_{\rm E}$. There is a strong correlation between the inner
disk temperature and the fraction of the disk component. Most of the
Comptonization-dominated ($>50\%$ total flux) spectra show $T_{\rm in} \sim 1$
keV with a high electron temperature of $>10$ keV, which may correspond to the
very high state in canonical black hole X-ray binaries (BHBs). By contrast, the
disk-dominated spectra have $T_{\rm in} \sim 2$ keV with a low temperature
($<10$ keV) and optically thick Comptonization, and show two separate branches
in the luminosity vs. innermost temperature ($L$ -- $T_{\rm in}$) diagram.
The lower branch clearly follows the $L \propto T_{\rm in}^4$-track.
Furthermore, applying the extended disk blackbody (or p-free disk) model, we find that 9
out of 12 datasets with disk luminosity above $0.3 L_{\rm E}$ prefer a flatter
temperature gradient than that in the standard disk ($p < 0.7$). We interpret
that, in the lower branch, the disk extends down to the innermost stable
circular orbit, and the source is most probably in the slim disk state. A
rapidly spinning black hole can explain both the lack of the
$L \propto T_{\rm in}^2$-track and a high value of spectral hardening factor
($\sim 4$) that would be required for a non-rotating black hole.
The spectra in the upper branch are consistent with
the picture of a truncated disk with low temperature Comptonization.
This state is uniquely observed from GRS 1915+105 among BHBs, which may
be present at near-Eddington luminosity.
\end{abstract}

\section{Introduction}
The dynamics and emission properties of supercritical (or super-Eddington)
accretion flows have been long-standing issues since the 1970s (Shakura and
Sunyaev 1973).
Theoretical studies on supercritical accretion are far from being complete
due to the complex interaction between radiation and matter,
despite recent progress in numerical simulations (e.g. Ohsuga et al.
2009).
From observational studies, supercritical accretion appears to be present
in some extragalactic objects, such as Narrow-line Seyfert 1 galaxies (NLS1s)
and Ultraluminous X-ray sources (ULXs) (Kato et al. 2008,
and references therein). 
In some NLS1s, the black hole mass can be estimated by means of reverberation
mapping (e.g. Wandel et al. 1999; Kaspi et al. 2000).
In the case of ULXs, however, the black hole mass is unknown and thus
remains controversial. Nevertheless, their luminosity is higher than the
Eddington luminosity of a stellar mass black hole.
Black hole mass is the key to initially identifying a supercritical
accreting system since the Eddington luminosity depends only on the black hole
mass.
In order to do a comprehensive study on supercritical accretion
processes, we need an object whose black hole mass can be well-constrained.
In addition, we also need adequate observational data.

One of the best objects for the study of supercritical accretion is the
Galactic microquasar, GRS 1915+105.
GRS 1915+105 has been active since its discovery
by the WATCH all-sky X-ray monitor on {\it GRANAT} (\cite{8}; 1994)
on 1992 August 15.
Its distance and inclination angle have been estimated from its super-luminal
jets (\cite{19}) to be $12.5 \pm 1.5$ kpc (see also \cite{7}; \cite{11}) and
$70^{\circ} \pm 2^{\circ}$, respectively.
Dozens of types of peculiar variability in the X-rays have been observed
(see \cite{6}; Klein-Wolt et al. 2002; Hannikainen et al. 2005),
most of which have never been observed in other Galactic X-ray
sources, including other known microquasars.
GRS 1915+105 is thus known to be a peculiar X-ray source in our Galaxy.
The jet feature and the variability of GRS 1915+105 are believed
to be an important key to studying the coupling between jets and accretion
disks in general (e.g. \cite{24}; \cite{23}).

While optical observations are severely hampered since GRS 1915+105 is
located close to the Galactic plane at $l = 45.37$, $b = -0.22$ (\cite{9}),
observations at other wavelengths prove to be quite fruitful.
From infrared spectroscopy, GRS 1915+105 was suggested to be
a low-mass X-ray binary (\cite{10}).
This result is further supported by Greiner et al. (2001a) and they also
suggested that the donor star is a K -- M giant with a narrow mass range
$\sim 1.0$ -- 1.5$M_{\odot}$.
The dynamical mass estimation results in a compact object
of $14 \pm 4 M_{\odot}$ (\cite{13}).
This classifies GRS 1915+105 as the most massive black hole X-ray binaries
(BHBs) in our Galaxy (for extragalactic BHBs, see, e.g. \cite{14};
\cite{15}). 

Since the dynamical mass can be estimated, GRS 1915+105 is known to
prefer being in a state where the luminosity is close to or even exceeding
its Eddington luminosity (\cite{20}), which corresponds to the supercritical
accretion process.
Provided that GRS 1915+105 does really exhibit supercritical accretion,
it can be used as a guide for the interpretation of other supercritical
accreting objects that lack sufficient data for a comprehensive study.

We use archival {\it RXTE} data mainly from 1999 to 2000 and apply the so-called
disk-corona model to the extracted spectra.
Our focus on the study of the supercritical accretion in GRS 1915+105
distinguishes this work from other previous extensive studies
(e.g. \cite{20}; \cite{81}; \cite{21}).
We also need to stress that we do not attempt to explain the nature of
GRS 1915+105 with its richness of variability on timescales from years down
to milliseconds (see, e.g. \cite{16}; \cite{6}) and its non-canonical soft
and hard states.
Those studies have been done by many authors (see, e.g. \cite{4}, b)
and the most recent work can be found in Rodriguez et al. (2008a, b). 
On the contrary, we choose data from when the source is more or less stable,
without strong variability features as classified in Belloni et al. (2000).
The constraint on the mass of GRS 1915+105 from dynamical estimates helps
us to study the detailed relation between the X-ray spectra and mass
accretion rate in terms of its luminosity during its supercritical accretion
state.

The layout of this paper is as follows:
We describe our data sample in section 2, and continue with describing the
model and the fitting analysis in section 3 and 4, respectively. 
Section 5 is devoted to the discussion, and finally section 6 concludes the
paper.
Throughout our study we adopt the following values: a distance of 12 kpc,
an inclination angle of 70$^{\circ}$, and a black hole mass of 14$M_{\odot}$.

\section{Data}
\begin{figure}
  \begin{center}
\centerline{\epsfig{file=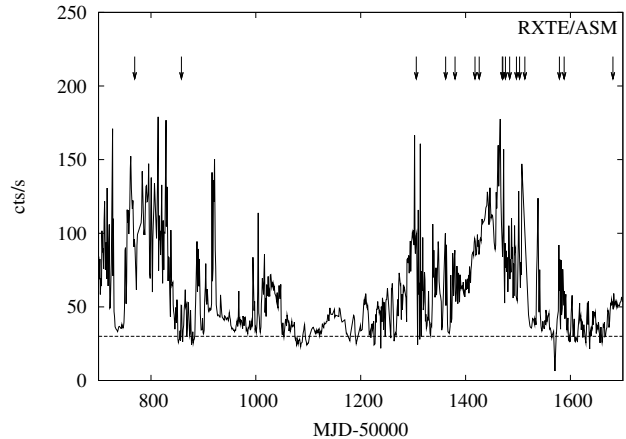,width=6.0cm,angle=270}}
  \end{center}
  \caption{{\it RXTE}/ASM lightcurve. The minimum count for soft state detection,
30 cts/s, is marked with the dotted line. The arrows mark the datasets of our final
sample (see text).}\label{Figure:2}
\end{figure}

\begin{table*}
  \caption{The final sample.
}\label{tab:first}
  \begin{center}
    \begin{tabular}{llccrr} 
\hline \hline

  Dataset & Observation ID & Observation Date & Observation Time & Duration (s) & Class (State)\\

  \hline
 1  &  20402-01-55-00 & 1997-11-17 & 05:27:06.6 & 15378 & $\delta$ (B,(A))\\
 2  &  30703-01-08-00 & 1998-02-14 & 23:51:55.3 & 4868 & $\phi$ (A)\\
 3  &  40703-01-14-01 & 1999-05-08 & 04:33:18.7 & 4319 & $\delta$ (B,(A))\\
 4  &  40703-01-14-02 & 1999-05-08 & 06:09:19 & 3085 & $\delta$ (B,(A))\\
 5  &  40703-01-20-02 & 1999-07-03 & 01:51:37.7 & 4190 & $\chi$ (C)\\
 6  &  40703-01-22-01 & 1999-07-21 & 01:57:09.5 & 2699 & $\chi$ (C)\\
 7  &  40703-01-28-00 & 1999-08-28 & 18:30:00.6 & 2891 & $\delta$ (B,(A))\\
 8  &  40703-01-28-01 & 1999-08-28 & 20:17:11.3 & 2220 & $\delta$ (B,(A))\\
 9  &  40703-01-29-00 & 1999-09-05 & 18:12:41.2 & 3210 & $\delta$ (B,(A))\\
 10 &  40703-01-29-01 & 1999-09-05 & 19:59:11.3 & 2520 & $\delta$ (B,(A))\\
 11 &  40115-01-08-01 & 1999-10-19 & 11:55:03.6 & 3514 & $\delta$ (B,(A))\\
 12 &  40703-01-34-00 & 1999-10-20 & 15:19:06.5 & 2731 & $\delta$ (B,(A))\\
 13 &  40703-01-34-01 & 1999-10-20 & 17:09:35.1 & 1862 & $\delta$ (B,(A))\\
 14 &  40703-01-35-00 & 1999-10-25 & 15:07:08.5 & 3108 & $\chi$ (C)\\
 15 &  40703-01-36-00 & 1999-11-02 & 14:48:37.4 & 2415 & $\chi$ (C)\\
 16 &  40703-01-38-03 & 1999-11-15 & 14:09:02.3 & 3420 & $\chi$ (C)\\
 17 &  40703-01-38-02 & 1999-11-15 & 15:43:01.6 & 3540 & $\chi$ (C)\\
 18 &  40703-01-38-01 & 1999-11-15 & 17:13:01.6 & 3900 & $\chi$ (C)\\
 19 &  40703-01-39-00 & 1999-11-21 & 12:09:00.9 & 3901 & $\delta$ (B,(A))\\
 20 &  40703-01-40-00 & 1999-12-01 & 11:45:03.2 & 3900 & $\chi$ (C)\\
 21 &  40703-01-40-02 & 1999-12-01 & 14:56:22.3 & 2912 & $\chi$ (C)\\
 22 &  40703-01-40-03 & 1999-12-01 & 16:30:03.5 & 3240 & $\chi$ (C)\\
 23 &  40702-01-04-00 & 2000-02-05 & 05:58:34.5 & 3004 & $\chi$ (C)\\ 
 24 &  40403-01-10-00 & 2000-02-14 & 10:28:11.1 & 2677 & $\chi$ (C)\\
 25 &  50703-01-10-01 & 2000-05-17 & 03:15:10.7 & 2753 & $\chi$ (C)\\
   \hline 
\end{tabular}
  \end{center}
\end{table*}

For the purpose of the present study, we deliberately choose the data
when the source was bright and more or less stable, that is when no large
flux variations are observed. 
Therefore, we only choose data from classes $\phi$, $\delta$,
and $\chi$ from the 12 known classes of X-ray variability in GRS 1915+105
(\cite{6}).
The variability selection is made by eye from the lightcurve
available through the {\it RXTE} standard products.

GRS 1915+105 has been monitored by {\it RXTE} since 1996,
once a week on average.
We select the {\it RXTE} data of GRS 1915+105 between January 1999 and May 2000.
This range of observation times contains data from epochs 3B, 4 and 5, and
consists of a total of 216 datasets to be examined.
Focusing on data with small variability, we manage to keep 23 datasets for further
fitting analysis.
In addition, we also analyze two disk-dominated datasets from 1997 (epoch
3A) and 1998 (epoch 3B), previously studied by Done et al. (2004).
To sum up, our final sample consists of 25 datasets (epoch 3A, 3B, 4, and 5).
The details of the selected datasets are given in Table 1.

We follow the standard data reduction procedure for {\it RXTE} data.
We use PCA data only for the fitting energy range we are interested in,
3 -- 25 keV, with 16-s time-resolution of the Standard 2 data. 
The lightcurves and the spectra of the source and background were extracted
using {\tt SAEXTRCT} version 4.2e. 
The latest version of {\tt RUNPCABACKEST} (version 4.0) was used to estimate
the background. The response files were created by using {\tt PCARSP} version
10.1.
The good time intervals are selected using the following conditions:
elevation angle greater than 10 degrees, the OFFSET less than 0.02,
and we use all detectors which were in operation during the observation time.
Dead-time correction is applied and 1.5\% systematic error is included
in each spectral bin.
We find no significant change in the fitting parameters when a 1\%
systematic error is applied.
Figure 1 shows the one-day average {\it RXTE}/ASM lightcurve of GRS 1915+105
(sum band intensity) from MJD 50700 -- 51700.
The arrows mark the datasets of our final sample.

We follow Belloni et al. (2000) for the description of states A, B, and C
in Table 1. States A and B correspond to disk-dominated spectra while state C
corresponds to the power-law type spectra. State A has lower flux and disk
temperature ($\sim 1$ keV) compared to those of state B. State B is associated
with high mass accretion rate and its spectra resemble that of the very high
state (VHS) in normal BHBs (e.g. Belloni et al. 2000, and
the references therein). State C is mostly found in variability class $\chi$,
while states A and B can be found in variability class $\delta$. Variability
class $\phi$ mostly consists of state A.

\section{Fitting Models}
In this section we describe the models used in our analysis.
We fit all the data (3 -- 25 keV) with an absorbed disk-corona model
employing two different disk models, one for the whole dataset and one applied
only to a subset. Details are as follow:

\subsection{Absorption model}
We model the absorption by the interstellar (as well as circumstellar) medium
by using the Tuebingen-Boulder interstellar medium absorption model
(\lq{\tt tbvarab}\rq{} in {\tt XSPEC11}).
The constraint on the elemental abundance in the direction of GRS 1915+105 is
adopted from the recent study by Ueda et al. (2009).
By using data from {\it Chandra} and {\it RXTE}, they make a robust
constraint of the interstellar (plus circumstellar) absorption components
towards GRS 1915+105.
This is, in a sense, an improvement in the study of GRS 1915+105 since
the absorption model significantly affects the energy range of the emission
from a black hole accretion disk.
The hydrogen column density is fixed at $2.78 \times 10^{22}$ cm$^{-2}$
with metal abundances larger than solar assumed.
This value is smaller than the commonly adopted value for this source in which
solar abundances are assumed, $4.7 \times 10^{22}$ cm$^{-2}$ (\cite{7}).

\subsection{Disk models}
We fit the disk component of the whole dataset with a multicolor disk
(MCD) model (\cite{41}; \cite{42}), a mathematical  approximation
model of the standard accretion disk (\cite{43}).

Given that some of our data have a luminosity close to or even
above the Eddington luminosity, we also fit a subset of the whole dataset
with another disk model, the extended disk blackbody
(extended DBB) model (the so-called $p$-free disk model; \cite{44}). This subset
contains data with disk-dominated spectra (see section 4.1.2 for details).

The standard theory of accretion disks gives an effective
temperature profile which is proportional to $r^{-3/4}$,
where $r$ is the disk radius.
As the luminosity approaches the Eddington luminosity, the standard
accretion disk model breaks down and should be replaced by the supercritical
accretion model (e.g. the slim disk model).
In the supercritical accretion regime, the temperature
profile is expected to become flatter at small radii; this is caused by the advective
motion of the trapped radiation.
In the extended DBB model, the effective temperature profile is assumed to
be $T_{\rm eff} \propto r^{-p}$, where $p$ is now a fitting parameter.
In principle, by using the extended DBB model, we can distinguish the slim disk
from that of the standard disk by spectral fitting (e.g. Vierdayanti
et al. 2006; Okajima et al. 2006; Tsunoda et al. 2006).
Furthermore, mass loss due to radiation pressure-driven
outflow is also known to modify the temperature profile (Shakura and Sunyaev
1973; Poutanen et al. 2007; Takeuchi et al. 2009). For a mass accretion rate of
$\dot{M} \propto r^{s}$, where $s$ is a positive parameter, the effective
temperature profile follows $T_{\rm eff} \propto r^{-(3-s)/4}$.
\begin{figure*}
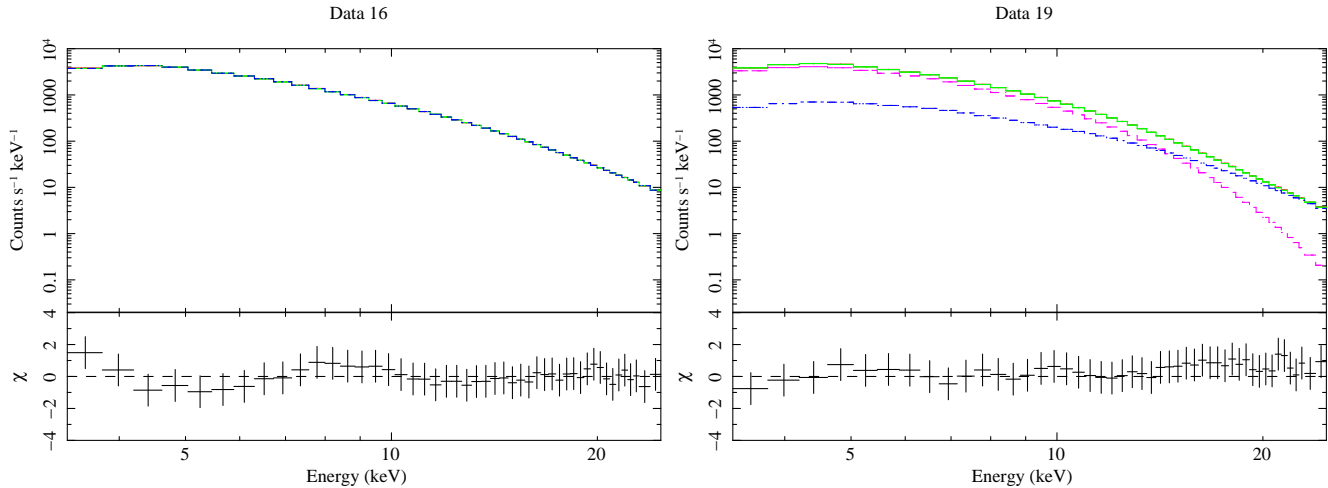

  \begin{center}
\centerline{\epsfig{file=fig2c.ps,width=6.5cm,angle=270}
 \epsfig{file=fig2d.ps,width=6.5cm,angle=270}}
  \end{center}
  \caption{The spectral fits and residuals of some representative data: dataset 16 (left) and dataset 19 (right). The total model is shown in green, while the disk and Comptonization components are shown in magenta and blue, respectively.}\label{Figure:2}
\end{figure*}

\subsection{Thermal Comptonization model}
We add a thermal Comptonization model to the disk models mentioned above
(\lq{\tt nthcomp}\rq{} in {\tt XSPEC11}; \cite{45}) to fit
the high energy part of the spectrum ($> 10$keV), that is, the coronal
component.
Most of our data prefer low electron temperature values ($<10$ keV) and
{\tt nthcomp} is one of the better Comptonization models for such cases. It is
also a consistent model for our present study since the seed photons can have
a disk spectrum (compare, for example, with \lq{\tt comptt}\rq{} in
{\tt XSPEC11} in which the seed photons are assumed to have a Wien spectrum).
The temperature of the seed photons of the corona is linked to the disk
temperature, by assuming that the seed photons are coming from the disk.

To make the fitting realistic, we also consider the reflection of up-scattered
photons by the disk based on the code by Magdziarz \& Zdziarski (1995)
({\tt pexriv} in {\tt XSPEC11}).
Here, however, we have to apply several assumptions.
We assume that the disk emissivity profile is proportional to 
$r^{-3}$, where $r$ is the disk radius.
Other important parameters for the reflection component are the reflection
amplification ($\equiv \Omega/2 \pi$), the ionization parameter, $\xi$
($\equiv L/nr^2$, where $L$ is the luminosity and $n$ is density),
and the inner and outer radius of the reflecting area of the disk, $R_{\rm in}$
and $R_{\rm out}$, respectively. The iron-K emission line is included in the model.

Throughout our study we treated these reflection parameters as free parameters
except for $R_{\rm out}$ that is always fixed at $10^5 r_{\rm g}$ ($r_{\rm g} \equiv GM/c^2$).
We find that most of our data prefer low $R_{\rm in}$ ($<100r_{\rm g}$),
inferring significant reflection from the inner part of the disk
although its error is large.
We thus fix $R_{\rm in}$ at 40$r_{\rm g}$, adopting the typical value from
the previous study by Done et al. (2004).
Low values of $\xi$ are also preferred by our data.
Therefore, we adopt the value of $\xi$ as found in Ueda et al. (2009);
that is, we fix $\xi$ at 40.
The overall fitting results, however, do not sensitively depend on our choice
of $\xi$.
Finally, we treat $\Omega/2 \pi$ as the only free parameter of the reflection
component. We, however, limit the maximum value of $\Omega/2 \pi$ to 1.

Several studies have reported the discovery of absorption line features
from this source (see, e.g. Kotani et al. 2000).
We find that the fitting residuals improved significantly when we include
an iron edge (modeled with \lq{\tt edge}\rq{} in {\tt XSPEC11}) to account
for absorption features of Fe XXVI ions as done in Ueda et al. (2009).
We thus expect that H-like absorption line of iron to be present in the
spectra of our final sample.
Therefore, we include an absorption line at 7.0 keV (K$\alpha$).
Regarding the absorption lines, we use a negative Gaussian model.
We fix the line width at 40 eV but left the normalization of this model
as a free parameter.
In addition, this normalization is also linked to the absorption depth of the
iron to make a more physical fitting.
This is necessary because both the equivalent width of an absorption line
and the optical depth of an edge are proportional to the normalization
of the Gaussian, provided that the absorption equivalent width is in a linear
part of the curve of growth (see Kotani et al. 2000).
\begin{figure}
  \begin{center}
\centerline{\epsfig{file=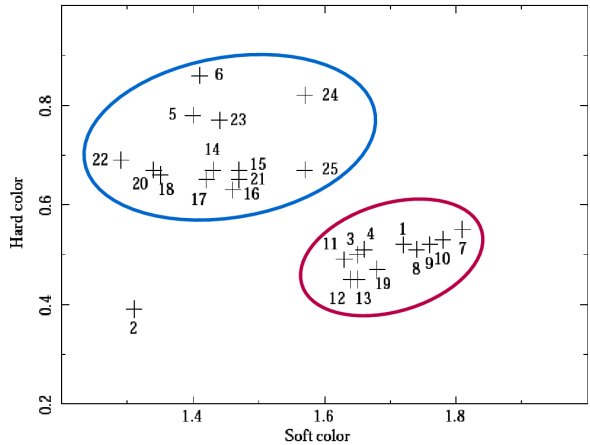,width=8.2cm,angle=0}}
  \end{center}
  \caption{Color-color diagram of our sample. The numbers represent the order of the observation time listed in table 1. Most data inside the blue circle have Comptonization-dominated spectra and those inside the red circle have disk-dominated spectra.}\label{Figure:1}
\end{figure}

\section{Fitting Results and Analysis}
As mentioned earlier, the energy range for fitting is set to be
3 -- 25 keV (PCA data only) and there are 25 datasets in our final sample (see
table 1). We first fit all the datasets with the (MCD+nthcomp) model.
See Figure 2 for two representative samples of disk-dominated spectra (right
panel) and Comptonization-dominated spectra (left panel).

A color-color diagram of our data sample is given in figure 3.
We follow Done and Gierli\'{n}ski (2003) in defining the soft and hard colors.
That is, the soft color is defined as the ratio of unabsorbed fluxes in the
4 -- 6.4 and 3 -- 4 keV bands, while the hard color as the ratio of unabsorbed
fluxes in the 9.7 -- 16 and 6.4 -- 9.7 keV bands.
As mentioned in section 2, we also include two datasets which were previously
studied by Done et al. (2004). Datasets 1 and 2 in this work correspond to
datasets 6 and 7 of Done et al. (2004).
The upper left group in figure 3 consists of data with
Comptonization-dominated spectra with one exception, dataset 25. 
On the other hand, the lower right group mostly consists of data with
disk-dominated spectra with two exceptions, datasets 11 and 12.  
Dataset 2 is Comptonization-dominated when fitted with our model as
opposed to the previous study. 

A second disk model (the extended DBB model) was applied to the subset in the
lower right of the color-color diagram (see figure 3, red circle) and dataset
25 from the upper left group.
In total, we fit 12 datasets with the (extended DBB model+nthcomp) model.
We obtain acceptable values of reduced chi-squared,
$\chi^2/$d.o.f. where d.o.f. means degrees of freedom, for overall fitting
(except dataset 2, where $\chi^2/$d.o.f.$>1.5$).
In fact, in most cases, the $\chi^2/$d.o.f. is rather low.
We found that this is caused by the moderately large systematic error value
(1.5\%) included in the fitting.
The results from fitting all the data with the (MCD+nthcomp) model are
given in tables 2 and 3, while the results from the (extended DBB+nthcomp)
model are given in table 4.
The unfolded spectra of all datasets in our final sample, together with the
corresponding components of the models, are given in figures 4 and 5
for the (MCD+nthcomp) model and figure 6 for the (extended DBB+nthcomp) model.

%
\begin{figure*}
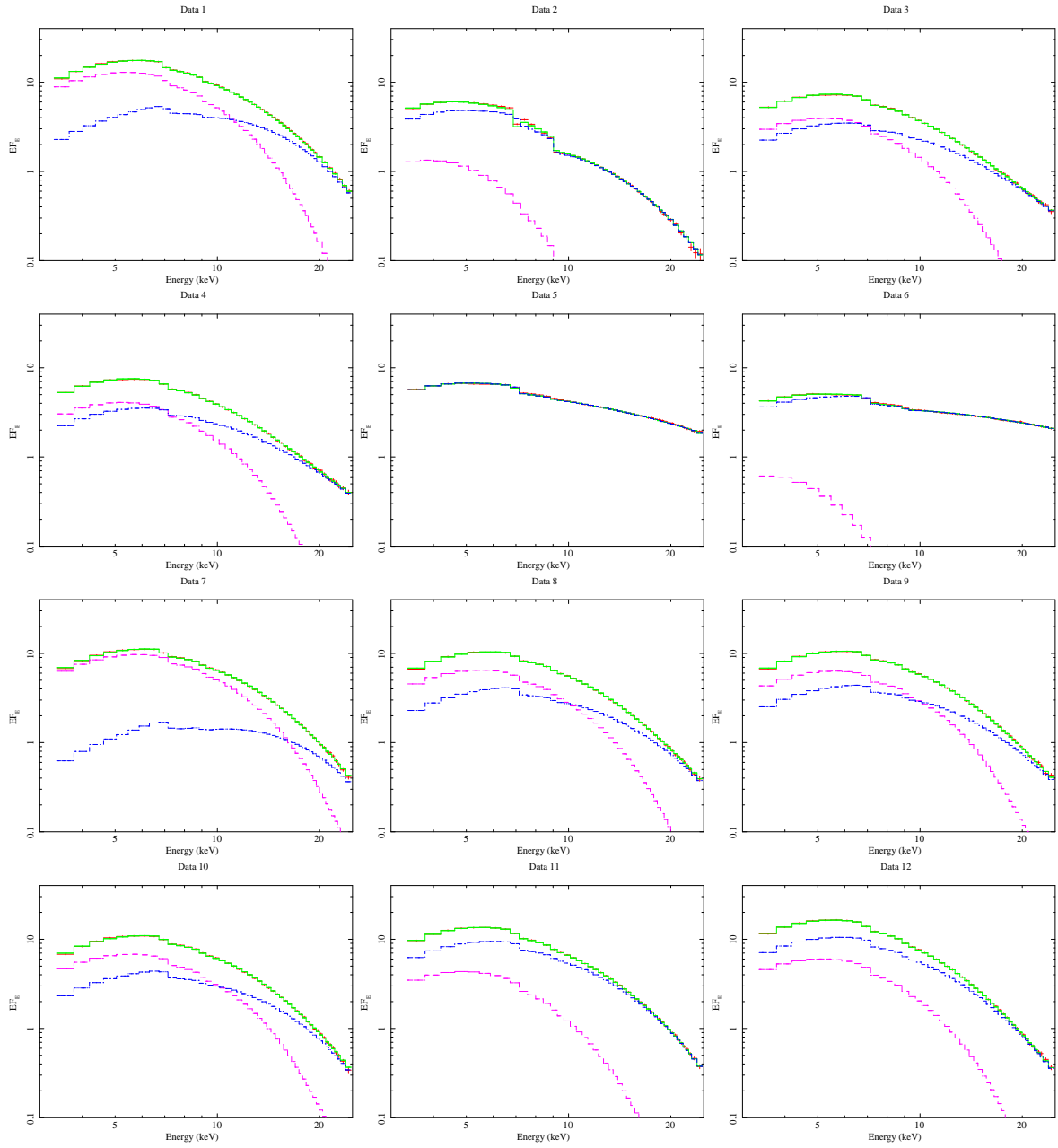

  \begin{center}
\centerline{\epsfig{file=fig4a.ps,width=4.2cm,angle=270}
 \epsfig{file=fig4b.ps,width=4.2cm,angle=270}
 \epsfig{file=fig4c.ps,width=4.2cm,angle=270}}
\centerline{\epsfig{file=fig4d.ps,width=4.2cm,angle=270}
 \epsfig{file=fig4e.ps,width=4.2cm,angle=270}
 \epsfig{file=fig4f.ps,width=4.2cm,angle=270}}
\centerline{\epsfig{file=fig4g.ps,width=4.2cm,angle=270}
 \epsfig{file=fig4h.ps,width=4.2cm,angle=270}
 \epsfig{file=fig4i.ps,width=4.2cm,angle=270}}
\centerline{\epsfig{file=fig4j.ps,width=4.2cm,angle=270}
 \epsfig{file=fig4k.ps,width=4.2cm,angle=270}
 \epsfig{file=fig4l.ps,width=4.2cm,angle=270}}
  \end{center}
  \caption{Unfolded spectra and model of datasets 1 -- 12.
The total, disk (MCD), and thermal Comptonization
components are shown in green, magenta, and blue lines, respectively.
The data, shown with red crosses, coincide with the total model.
}\label{Figure:4}
\end{figure*}

\begin{figure*}
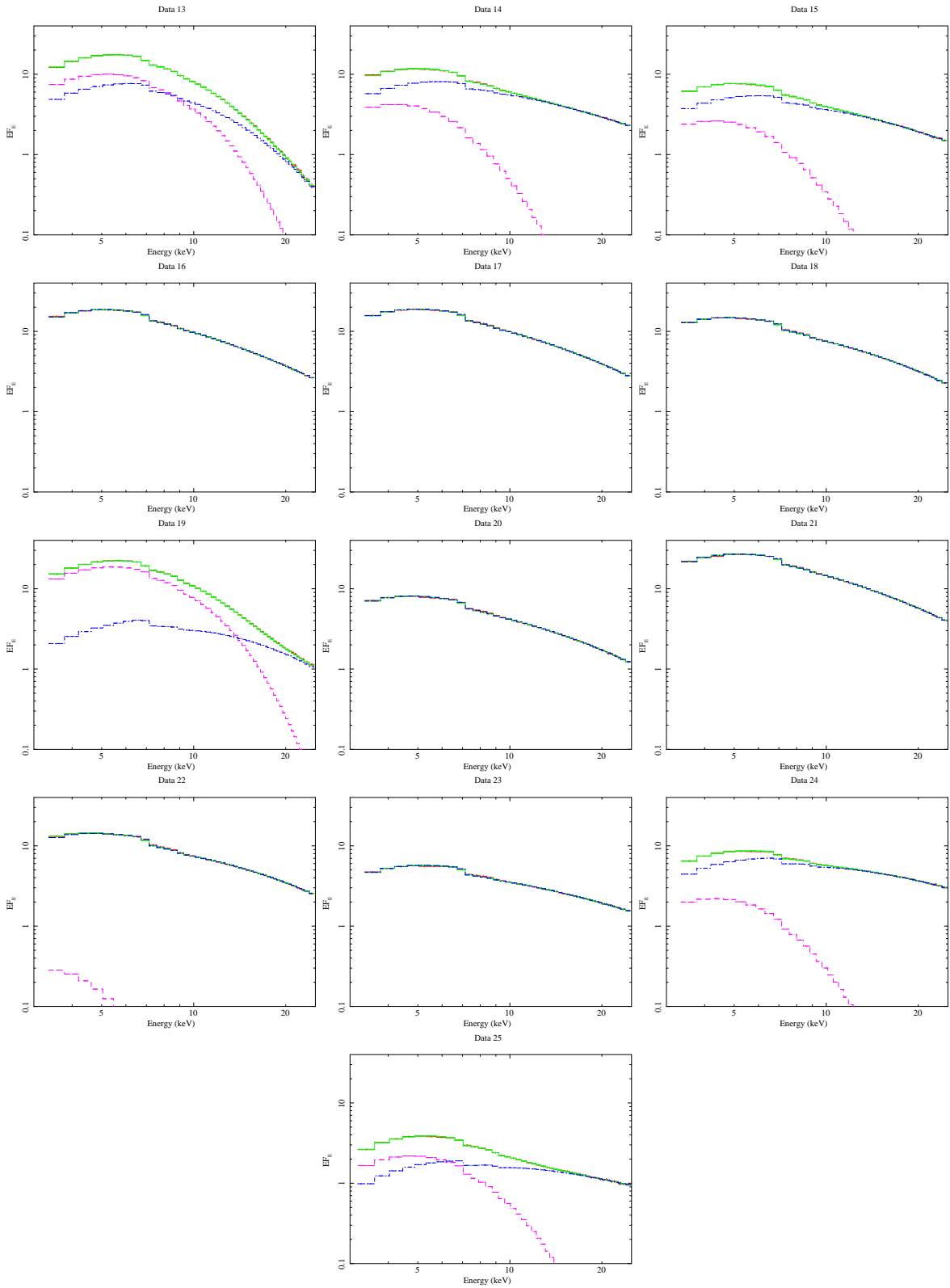

  \begin{center}
\centerline{\epsfig{file=fig5m.ps,width=4.2cm,angle=270}
 \epsfig{file=fig5n.ps,width=4.2cm,angle=270}
 \epsfig{file=fig5o.ps,width=4.2cm,angle=270}}
\centerline{\epsfig{file=fig5p.ps,width=4.2cm,angle=270}
 \epsfig{file=fig5q.ps,width=4.2cm,angle=270}
 \epsfig{file=fig5r.ps,width=4.2cm,angle=270}}
\centerline{\epsfig{file=fig5s.ps,width=4.2cm,angle=270}
 \epsfig{file=fig5t.ps,width=4.2cm,angle=270}
 \epsfig{file=fig5u.ps,width=4.2cm,angle=270}}
\centerline{\epsfig{file=fig5v.ps,width=4.2cm,angle=270}
 \epsfig{file=fig5w.ps,width=4.2cm,angle=270}
 \epsfig{file=fig5x.ps,width=4.2cm,angle=270}}
\centerline{\epsfig{file=fig5y.ps,width=4.2cm,angle=270}}
  \end{center}
  \caption{Same as figure 4 for datasets 13 -- 25.
}\label{Figure:5}
\end{figure*}

\subsection{Disk Components}
\begin{table*}
  \caption{MCD components of the fitting. The unit for $kT_{\rm in}$ is keV. $r_{\rm in}$ is not calculated from the MCD model normalization (see text). $F_{\rm disk}$ and $F_{\rm disk}^{p}$ are the 0.01 -- 100 flux ($10^{-8}$ erg cm$^{-2}$ s$^{-1}$) and the photon flux (in photon s$^{-1}$ cm$^{-2}$) of the disk component, respectively. \% Disk is the disk fraction on the spectrum in per cent. The values of $F_{\rm disk}$, $F_{\rm disk}^{p}$, \% Disk, and $L_{\rm disk}/L_{\rm E}$ which are less than $10^{-5}$ are written as 0. 
}\label{tab:third}
  \begin{center}
    \begin{tabular}{lcrrrrrr}
\hline \hline
& & & & & & & \\
 Dataset  & $kT_{\rm in}$ & $r_{\rm in}/r_{\rm S}$  & $F_{\rm disk}$ & $F_{\rm disk}^{p}$ & \% Disk & $L_{\rm disk}/L_{\rm E}$ & $\chi^{2}$/d.o.f \\[1ex]
  \hline
& & & & & & & \\
$1$  & $1.81_{-0.06}^{+0.1}$ & $0.50 \pm 0.07$ & $5.21$ & $22.8$ & $74.6$ & $0.68 \pm 0.20$ & $0.22$ \\ [2ex]
$2$  & $1.06_{-0.09}^{+0.1}$ & $1.80 \pm 0.25$ & $0.95$ & $6.77$ & $29.0$ & $0.12 \pm 0.04$ & $1.57$ \\ [2ex]
$3$  & $1.71_{-0.03}^{+0.03}$ & $0.57 \pm 0.08$ & $1.63$ & $7.52$ & $53.0$ & $0.21 \pm 0.06$ & $0.46$ \\ [2ex]
$4$  & $1.74_{-0.02}^{+0.03}$ & $0.55 \pm 0.07$ & $1.68$ & $7.61$ & $53.5$ & $0.22 \pm 0.06$ & $0.37$ \\ [2ex]
$5$  & $0.84_{-0.05}^{+0.08}$ & $3.45 \pm 0.50$ & $0.001$ & $0.009$ & $0.02$ & 1E-4 $\pm$ 4E-5 & $0.52$ \\ [2ex]
$6$  & $0.89_{-0.15}^{+0.07}$ & $2.42 \pm 0.35$ & $0.47$ & $3.98$ & $13.7$ & $0.06 \pm 0.02$ & $0.43$ \\ [2ex]
$7$  & $2.05_{-0.03}^{+0.02}$ & $0.21 \pm 0.03$ & $3.70$ & $14.4$ & $87.3$ & $0.48 \pm 0.14$ & $0.42$ \\ [2ex]
$8$  & $1.87_{-0.03}^{+0.04}$ & $0.49 \pm 0.07$ & $2.57$ & $10.9$ & $62.8$ & $0.33 \pm 0.10$ & $0.28$ \\ [2ex]
$9$  & $1.93_{-0.03}^{+0.04}$ & $0.49 \pm 0.07$ & $2.46$ & $10.1$ & $59.5$ & $0.32 \pm 0.10$ & $0.51$ \\ [2ex]
$10$ & $1.92_{-0.03}^{+0.04}$ & $0.46 \pm 0.06$ & $2.66$ & $11.0$ & $63.0$ & $0.35 \pm 0.10$ & $0.57$ \\ [2ex]
$11$ & $1.57_{-0.04}^{+0.05}$ & $1.09 \pm 0.15$ & $1.87$ & $9.52$ & $32.8$ & $0.24 \pm 0.07$ & $0.20$ \\ [2ex]
$12$ & $1.68_{-0.03}^{+0.04}$ & $1.06 \pm 0.15$ & $2.48$ & $11.7$ & $36.7$ & $0.32 \pm 0.09$ & $0.30$ \\ [2ex]
$13$ & $1.73_{-0.03}^{+0.03}$ & $0.82 \pm 0.11$ & $4.09$ & $18.6$ & $57.3$ & $0.53 \pm 0.15$ & $0.32$ \\ [2ex]
$14$ & $1.23_{-0.03}^{+0.03}$ & $1.64 \pm 0.24$ & $2.21$ & $13.9$ & $34.9$ & $0.29 \pm 0.08$ & $0.25$ \\ [2ex]
$15$ & $1.26_{-0.03}^{+0.03}$ & $1.26 \pm 0.18$ & $1.34$ & $8.27$ & $33.5$ & $0.17 \pm 0.05$ & $0.24$ \\ [2ex]
$16$ & $1.10_{-0.03}^{+0.08}$ & $3.37 \pm 0.49$ & $0$   & $0$ & $0$    & $0$    & $0.29$ \\ [2ex]
$17$ & $1.00_{-0.02}^{+0.12}$ & $4.12 \pm 0.59$ & $0$   & $0$ & $0$    & $0$    & $0.29$ \\ [2ex]
$18$ & $0.87_{-0.04}^{+0.04}$ & $4.93 \pm 0.71$ & $0$ & $0$ & $0$ & $0$ & $0.24$ \\ [2ex]
$19$ & $1.84_{-0.03}^{+0.03}$ & $0.48 \pm 0.07$ & $7.43$ & $32.1$ & $82.3$ & $0.97 \pm 0.28$ & $0.43$ \\ [2ex]
$20$ & $0.86_{-0.04}^{+0.04}$ & $3.76 \pm 0.54$ & $0$   & $0$ & $0$    & $0$    & $0.38$ \\ [2ex]
$21$ & $1.08_{-0.02}^{+0.11}$ & $4.14 \pm 0.60$ & $0$& $0$ & $0$ & $0$& $0.37$ \\ [2ex]
$22$ & $0.76_{-0.08}^{+0.09}$ & $6.61 \pm 0.95$ & $0.28$ & $2.81$ & $2.85$ & $0.04 \pm 0.01$ & $0.26$ \\ [2ex]
$23$ & $0.90_{-0.04}^{+0.11}$ & $2.72 \pm 0.39$ & $0$   & $0$ & $0$    & $0$    & $0.52$ \\ [2ex]
$24$ & $1.27_{-0.05}^{+0.05}$ & $1.33 \pm 0.19$ & $1.12$ & $6.82$ & $24$ & $0.15 \pm 0.04$ & $0.40$ \\ [2ex]
$25$ & $1.51_{-0.03}^{+0.03}$ & $0.49 \pm 0.07$ & $0.98$ & $5.1$ & $51.1$ & $0.13 \pm 0.04$ & $0.40$ \\ [1ex]
  \hline 
\end{tabular}
  \end{center}
\end{table*}

\begin{table*}
  \caption{Thermal Comptonization component of the fitting. The unit for $kT_{\rm e}$ is keV. $F_{\rm THC}$ and $F_{\rm THC}^{p}$ are the 0.01 -- 100 keV flux ($10^{-8}$ erg cm$^{-2}$ s$^{-1}$) and the photon flux (in photon s$^{-1}$ cm$^{-2}$) of the Comptonized component, respectively. $\tau$ is the scattering optical depth and $L$ is the total luminosity (see text).
}\label{tab:third}
  \begin{center}
    \begin{tabular}{lccrrcrr} 
\hline \hline
& & & & & & & \\
 Dataset  & $kT_{\rm e}$ & $\Gamma$  & $F_{\rm THC}$ & $F_{\rm THC}^{p}$ & $\tau_{\rm e}$ & $\Omega/2 \pi$ & $L/L_{\rm E}$ \\[1ex]
  \hline
& & & & & & & \\
$1$  & $3.08_{-0.06}^{+0.07}$ & $2.12_{-0.06}^{+0.06}$ & $1.77$ & $4.83$ & $9.00$ & $0.69$ & $0.76 \pm 0.22$ \\ [2ex]
$2$  & $3.34_{-0.50}^{+0.20}$ & $3.16_{-0.05}^{+0.05}$ & $2.31$ & $12.6$ & $5.09$ & $1.00$ & $0.23 \pm 0.06$ \\ [2ex]
$3$  & $10.6_{-1.57}^{+2.54}$ & $3.74_{-0.06}^{+0.07}$ & $1.44$ & $5.35$ & $1.88$ & 4E-8 & $0.28 \pm 0.08$ \\ [2ex]
$4$  & $11.5_{-2.01}^{+3.58}$ & $3.75_{-0.07}^{+0.08}$ & $1.46$ & $5.29$ & $1.78$ & 2E-6 & $0.28 \pm 0.08$ \\ [2ex]
$5$  & $17.1_{-1.70}^{+2.32}$ & $2.76_{-0.01}^{+0.01}$ & $4.25$ & $23.0$ & $2.10$ & $0.30$ & $0.19 \pm 0.05$ \\ [2ex]
$6$  & $25.0_{-3.33}^{+0.0}$ & $2.58_{-0.01}^{+0.003}$ & $2.94$ & $13.4$ & $1.78$ & $0.41$ & $0.19 \pm 0.06$ \\ [2ex]
$7$  & $3.38_{-0.07}^{+0.08}$ & $1.94_{-0.03}^{+0.03}$ & $0.54$ & $1.22$ & $9.64$ & $1.00$ & $0.51 \pm 0.15$ \\ [2ex]
$8$  & $4.60_{-0.24}^{+0.25}$ & $3.27_{-0.06}^{+0.05}$ & $1.52$ & $5.10$ & $3.99$ & $0.41$ & $0.40 \pm 0.12$ \\ [2ex]
$9$  & $6.30_{-0.53}^{+0.75}$ & $3.81_{-0.08}^{+0.09}$ & $1.67$ & $5.71$ & $2.64$ & $0.30$ & $0.39 \pm 0.11$ \\ [2ex]
$10$  & $3.61_{-0.10}^{+0.11}$ & $2.95_{-0.04}^{+0.05}$ & $1.56$ & $5.00$ & $5.31$ & $0.49$ & $0.42 \pm 0.12$ \\ [2ex]
$11$ & $3.38_{-0.05}^{+0.07}$ & $3.09_{-0.02}^{+0.03}$ & $3.84$ & $15.0$ & $5.19$ & $0.21$ & $0.41 \pm 0.12$ \\ [2ex]
$12$ & $5.71_{-0.46}^{+0.49}$ & $4.40_{-0.09}^{+0.07}$ & $4.29$ & $17.4$ & $2.32$ & $0.27$ & $0.51 \pm 0.15$ \\ [2ex]
$13$ & $5.53_{-0.35}^{+0.56}$ & $3.98_{-0.07}^{+0.09}$ & $3.05$ & $11.5$ & $2.72$ & $0.37$ & $0.67 \pm 0.19$ \\ [2ex]
$14$ & $25_{-4.24}^{+0.0}$ & $2.89_{-0.02}^{+0.004}$ & $4.12$ & $16.4$ & $1.49$ & $0.12$ & $0.47 \pm 0.14$ \\ [2ex]
$15$ & $16.6_{-2.10}^{+2.93}$ & $2.86_{-0.02}^{+0.02}$ & $2.67$ & $10.5$ & $2.03$ & $0.21$ & $0.29 \pm 0.08$ \\ [2ex]
$16$ & $25_{-3.87}^{+0.0}$ & $3.33_{-0.02}^{+0.004}$ & $9.94$ & $49.8$ & $1.18$ & $0.24$ & $0.44 \pm 0.13$ \\ [2ex]
$17$ & $14.4_{-1.17}^{+1.55}$ & $3.15_{-0.01}^{+0.02}$ & $10.6$ & $56.2$ & $1.95$ & $0.27$ & $0.47 \pm 0.14$ \\ [2ex]
$18$ & $11.7_{-0.72}^{+0.85}$ & $3.04_{-0.01}^{+0.01}$ & $9.09$ & $53.3$ & $2.37$ & $0.31$ & $0.40 \pm 0.12$ \\ [2ex]
$19$ & $7.51_{-0.65}^{+0.69}$  & $2.84_{-0.04}^{+0.04}$ & $1.60$ & $4.77$ & $3.52$ & $0.51$ & $1.04 \pm 0.30$ \\ [2ex]
$20$ & $11.2_{-0.68}^{+0.75}$ & $3.03_{-0.01}^{+0.01}$ & $5.01$ & $29.7$ & $2.46$ & $0.32$ & $0.22 \pm 0.06$ \\ [2ex]
$21$ & $15.6_{-1.50}^{+1.95}$ & $3.20_{-0.02}^{+0.02}$ & $14.4$ & $71.6$ & $1.80$ & $0.27$ & $0.64 \pm 0.18$ \\ [2ex]
$22$ & $10.6_{-0.54}^{+0.69}$ & $2.91_{-0.01}^{+0.01}$ & $9.66$ & $62.1$ & $2.70$ & $0.42$ & $0.47 \pm 0.13$ \\ [2ex]
$23$ & $25_{-0.01}^{+0.0}$ & $2.85_{-3.08}^{+0.02}$ & $3.47$ & $17.9$ & $1.51$ & $0.29$ & $0.15 \pm 0.04$ \\ [2ex]
$24$ & $14.1_{-1.27}^{+1.62}$ & $2.52_{-0.01}^{+0.01}$ & $3.54$ & $12.0$ & $2.76$ & $0.12$ & $0.30 \pm 0.09$ \\ [2ex]
$25$ & $21.2_{-4.36}^{+3.81}$ & $2.50_{-0.02}^{+0.01}$ & $0.94$ & $2.10$ & $2.69$ & 4e-5 & $0.17 \pm 0.05$ \\ [1ex]
  \hline 
\end{tabular}
  \end{center}
\end{table*}
\begin{figure*}
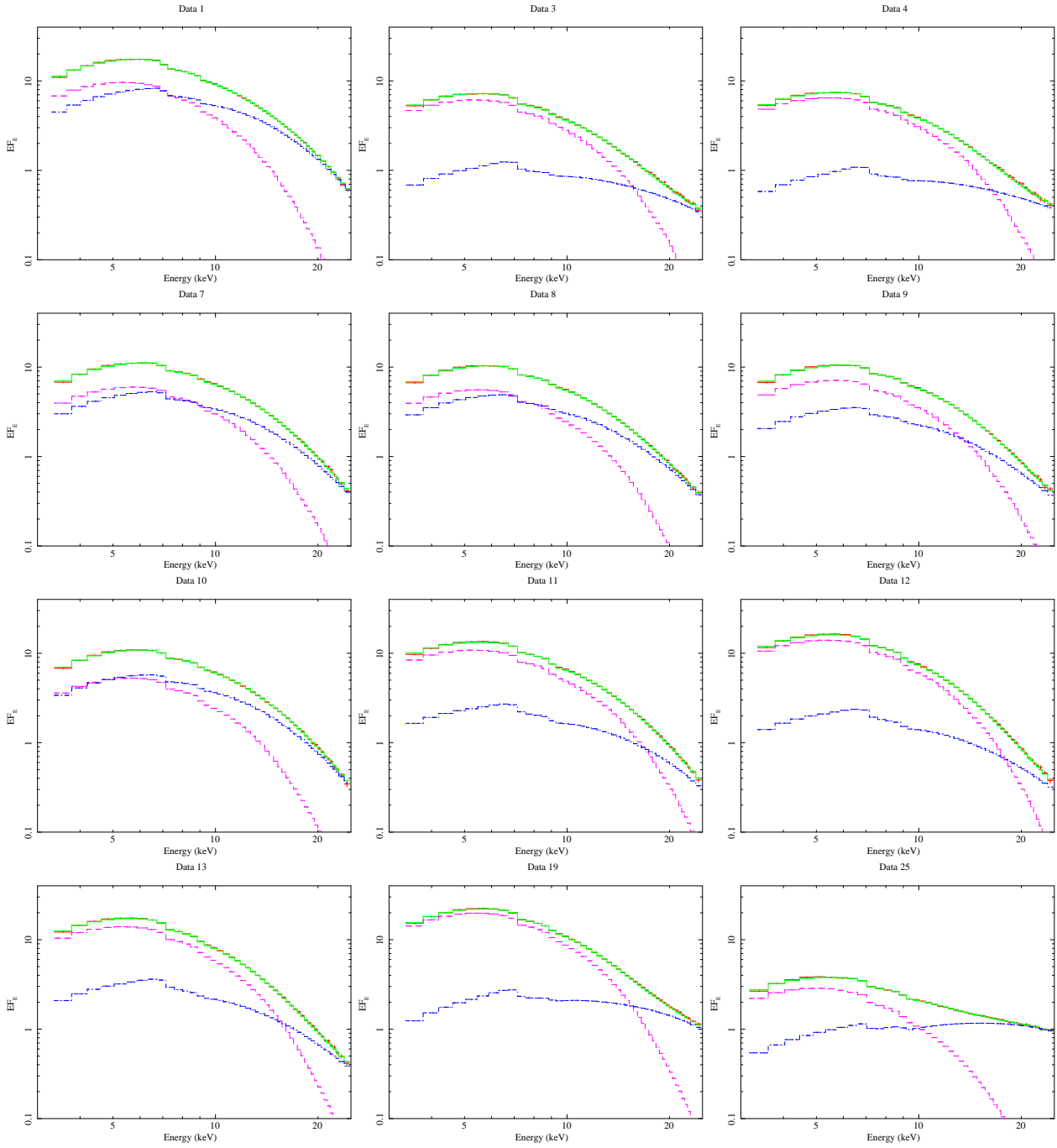

  \begin{center}
\centerline{\epsfig{file=fig6a.ps,width=4.5cm,angle=270}
 \epsfig{file=fig6b.ps,width=4.5cm,angle=270}
 \epsfig{file=fig6c.ps,width=4.5cm,angle=270}}
\centerline{\epsfig{file=fig6d.ps,width=4.5cm,angle=270}
 \epsfig{file=fig6e.ps,width=4.5cm,angle=270}
 \epsfig{file=fig6f.ps,width=4.5cm,angle=270}}
\centerline{\epsfig{file=fig6g.ps,width=4.5cm,angle=270}
 \epsfig{file=fig6h.ps,width=4.5cm,angle=270}
 \epsfig{file=fig6i.ps,width=4.5cm,angle=270}}
\centerline{\epsfig{file=fig6j.ps,width=4.5cm,angle=270}
 \epsfig{file=fig6k.ps,width=4.5cm,angle=270}
 \epsfig{file=fig6l.ps,width=4.5cm,angle=270}}
  \end{center}
  \caption{Unfolded spectra and model of the disk-dominated data. The total,
disk ($p$-free disk), and thermal Comptonization components are shown in green, magenta, and blue lines, respectively. The data, shown with red crosses, coincide with the total component. See text for details.
}\label{Figure:4}
\end{figure*}
\begin{table*}
  \caption{Fit results of the extended DBB plus thermal Comptonization model. The units for $kT_{\rm in}$, $kT_{\rm e}$, $L_{\rm disk}/L_{\rm E}$, and $L/L_{\rm E}$ are the same as in table 2 and 3.
}\label{tab:third}
  \begin{center}
    \begin{tabular}{lccccrrrr}
\hline \hline
& &\\
 Dataset  & $kT_{\rm in}$ & $p$  & $kT_{\rm e}$ & $\Gamma$ & \% Disk & $L_{\rm disk}/L_{\rm E}$ & $L/L_{\rm E}$ & $\chi^{2}$/d.o.f \\
[1ex]
  \hline
& & & & & & & \\
$1$  & $1.85_{-0.03}^{+0.03}$ & $0.69 \pm 0.12$ & $3.64 \pm 0.76$ & $2.93 \pm 2.00$ & $56.7$ & $0.55 \pm 0.16$ & $0.70 \pm 0.20$ & $0.45$ \\ [2ex]
$3$  & $2.06_{-0.01}^{+0.01}$ & $0.59 \pm 0.02$ & $8.09 \pm 1.86$ & $2.86 \pm 0.77$ & $85.0$ & $0.48 \pm 0.14$ & $0.51 \pm 0.15$ & $0.92$ \\ [2ex]
$4$  & $2.11_{-0.02}^{+0.02}$ & $0.59 \pm 0.02$ & $13.6 \pm 3.60$ & $2.93 \pm 0.76$ & $87.2$ & $0.49 \pm 0.14$ & $0.51 \pm 0.15$ & $0.79$ \\ [2ex]
$7$  & $2.03_{-0.02}^{+0.02}$ & $0.74 \pm 0.05$ & $8.28 \pm 1.86$ & $4.43 \pm 0.75$ & $54.2$ & $0.30 \pm 0.09$ & $0.39 \pm 0.11$ & $0.67$ \\ [2ex]
$8$  & $1.91_{-0.02}^{+0.02}$ & $0.72 \pm 0.04$ & $7.05 \pm 1.15$ & $4.05 \pm 0.05$ & $54.4$ & $0.30 \pm 0.09$ & $0.39 \pm 0.11$ & $0.64$ \\ [2ex]
$9$  & $2.06_{-0.02}^{+0.02}$ & $0.68 \pm 0.03$ & $25.0 \pm 0.0$ & $4.52 \pm 0.09$ & $67.9$ & $0.40 \pm 0.11$ & $0.46 \pm 0.13$ & $0.78$ \\ [2ex]
$10$ & $1.94_{-0.02}^{+0.02}$ & $0.75 \pm 0.06$ & $5.13 \pm 0.11$ & $3.94 \pm 0.05$ & $48.7$ & $0.27 \pm 0.08$ & $0.36\pm 0.10$ & $0.67$ \\ [2ex]
$11$ & $2.14_{-0.01}^{+0.03}$ & $0.56 \pm 0.01$ & $4.92 \pm 0.07$ & $3.15 \pm 0.03$ & $83.1$ & $1.06 \pm 0.30$ & $1.13 \pm 0.33$ & $1.03$ \\ [2ex]
$12$ & $2.03_{-0.15}^{+0.15}$ & $0.6 \pm 0.04$ & $9.34 \pm 6.07$ & $3.66 \pm 0.19$ & $87.6$ & $1.06 \pm 0.30$ & $1.11 \pm 0.32$ & $0.69$ \\ [2ex]
$13$ & $1.92_{-0.01}^{+0.05}$ & $0.65 \pm 0.07$ & $6.92 \pm 17.3$ & $3.61 \pm 4.33$ & $81.5$ & $0.89 \pm 0.25$ & $0.96 \pm 0.28$ & $0.76$ \\ [2ex]
$19$ & $1.92_{-0.14}^{+0.14}$ & $0.69 \pm 0.07$ & $6.00 \pm 5.19$ & $2.41 \pm 1.56$ & $88.9$ & $1.12 \pm 0.32$ & $1.17 \pm 0.34$ & $0.28$ \\ [2ex]
$25$ & $1.95_{-0.04}^{+0.12}$ & $0.55 \pm 0.05$ & $6.94 \pm 0.35$ & $1.91 \pm 0.10$ & $76.0$ & $0.33 \pm 0.10$ & $0.37 \pm 0.11$ & $0.60$ \\ [2ex]
  \hline 
\end{tabular}
  \end{center}
\end{table*}

\subsubsection{MCD components}
Our final sample has $kT_{\rm in}$ ranges between 0.7 -- 2 keV
while the total luminosity ranges from $\sim 0.15$ -- 1.04$L_{\rm E}$,
where $L_{\rm E}$ is the Eddington luminosity
($\equiv 1.5 \times 10^{38} (M/M_{\odot})$ erg s$^{-1}$).
Data with $kT_{\rm in}$ within 1.5 -- 2 keV are mostly
those with disk-dominated spectra (see figure 7, filled stars).
The inner disk temperature in our sample is too high for a 14$M_{\odot}$ 
non-rotating black hole.
A rotating black hole is, thus, often suggested to explain a too high
disk temperature phenomena in this source (McClintock et al. 2006; Middleton
et al. 2006; Zhang et al. 1997).
Caution needs to be taken for some Comptonization-dominated data. As shown
in table 2, some of these Comptonization-dominated data (i.e. datasets 16, 17,
18, 20, 21, and 23) have a very low disk fraction , $ \sim 0$. In this case,
the values of the disk temperature from the fitting become less reliable since
the disk contribution is negligible.

The disk inner radius can be derived from the normalization of the MCD model,
i.e., normalization $\equiv ((D/10 {\rm kpc})^2/(r_{\rm in} \ {\rm [km]})^2) \times \cos{i}$, where $D$ is the distance to the source in kpc and $i$ is the
inclination angle of the system. Note that $r_{\rm in}$ is the apparent inner
disk radius without relativistic ($\zeta$) and spectral hardening
factor ($\kappa$) corrections.
The true inner radius can be obtained by using the following formula:
$r_{\rm in}^{\rm true}=\kappa^{2}\zeta r_{\rm in}^{\rm apparent}$ 
(see Kubota et al.1998). That is, when we mention $r_{\rm in}$, we refer
to the $r_{\rm in}^{\rm true}$ throughout this paper unless stated otherwise.
However, since we use a disk-corona model, we have to take into account
the effect of the thermal Comptonization in the spectrum.
We calculate the apparent disk inner radius by using the formula (A1) shown in
Kubota and Makishima (2004) in which spherical geometry is assumed for the
thermal Comptonization emission.
However, here we assume that half the photons in the corona are injected back
into the accretion disk due to the large optical depth, as done in Ueda et al.
(2009).

By using this formula, the obtained true inner radius of the disk is mostly
within $\sim 0.5$ -- 7$r_{\rm S}$, where $r_{\rm S}$ is about 42 km for a
14$M_{\odot}$ black hole ($r_{\rm S} \equiv 2GM/c^2$ is a Schwarzschild
radius), assuming a non-rotating black hole. 
Here we have used $\kappa = 1.7$ and $\zeta=0.412$ to derive these
true inner radii. 
We find that the data whose disk fraction is larger than 50\% of the total
component have a low inner disk radius, $< 1r_{\rm S}$ (see figure 7,
filled stars). 
If, however, we choose higher $\kappa$, say $\kappa=3$ (note that in figure 10
the upper branch in the lower panel coincides with the line of $\kappa=3$
for a non-rotating black hole),
$r_{\rm in}$ would range within $\sim$ 1.5 -- 22$r_{\rm S}$,
that is $(3/1.7)^{2}$ times larger than that of $\kappa=1.7$ case.

We find a strong correlation between the disk parameters and the fraction of
the disk component.
As shown in figure 7, the inner disk temperature increases as the fraction
of the disk component increase.
The inner disk radius follows the opposite trend.
We attempt to interpret this correlation in section 5.

\subsubsection{Extended disk blackbody components}
For consistency, when the extended DBB model is used to fit
the disk component, we modify the temperature profile of the seed photons
in the thermal Comptonization model.
That is, the temperature profile is set to be $T_{\rm eff} \propto r^{-p}$,
with $p$ ranging from 0.5 to 0.75, instead of using the MCD temperature profile.

We only fit a group of data in the lower right of color-color diagram
(see figure 3, red circle) and dataset 25 from the upper left group
(blue circle). 
That is, we only choose data with disk-dominated
(or suspected, such as datasets 11 and 12) spectra.
We are interested in finding deviations from the standard
high/soft state spectra as the luminosity approaches or exceeds the
Eddington luminosity.
In total, we fit 12 datasets with the extended DBB model+modified thermal
Comptonization model.

The unfolded spectra and all components of the model are shown in
figure 6. Results of the fits are summarized in table 4.
As shown in table 4 (and figure 6), the disk fraction of the
twelve datasets is mostly larger than 75\%.
Moreover, disk fraction in datasets 11 and 12 increases significantly and
even exceeds the fraction of the thermal Comptonization component.

The radial temperature gradient, $p$, is shown in column 3 of table 4.
We can see that the $p$-values tend to deviate from 0.75.
In fact, nine out of twelve datasets (1, 3, 4, 9, 11, 12, 13, 19, and 25)
prefer a $p$-value which is less than 0.7 with less than 20\% error.
Moreover, four out of the nine datasets (3, 4, 11, 25) prefer a $p$-value
which is less than 0.6 with less than 10\% error.
The disk temperature of most datasets tends to increase compared to those of
the MCD+thermal Comptonization model, that is $kT_{\rm in} \sim 1.8$
-- 2.1 keV.
\begin{figure}
  \begin{center}
\centerline{\epsfig{file=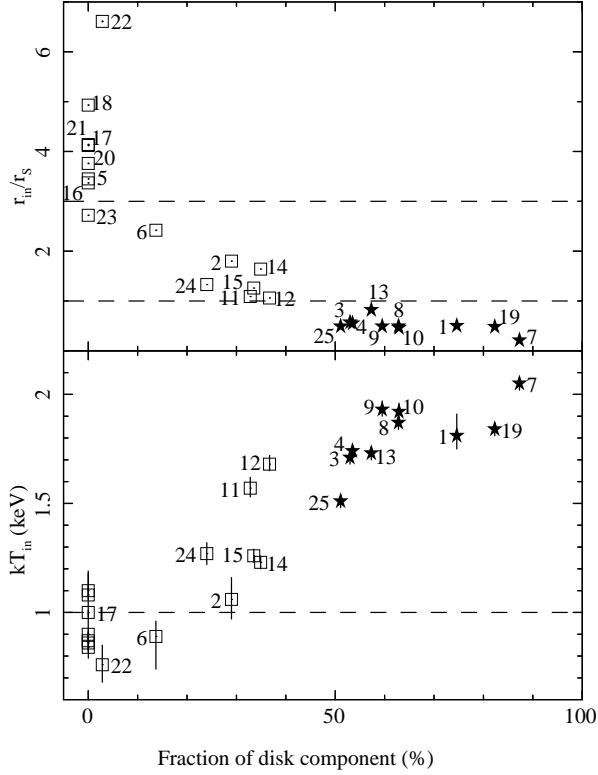,width=8.0cm,angle=0}}
  \end{center}
  \caption{$kT_{\rm in}$ and $r_{\rm in}/r_{\rm S}$ obtained from the fitting. The dotted line in the lower panel shows $kT_{\rm in}=1$ keV. The dotted lines in the upper panel mark $r_{\rm in}/r_{\rm S}=$1 and 3. Comptonization-dominated data are shown as open squares while disk-dominated data are shown as filled stars.}\label{Figure:4}
\end{figure}
\begin{figure}
  \begin{center}
\centerline{\epsfig{file=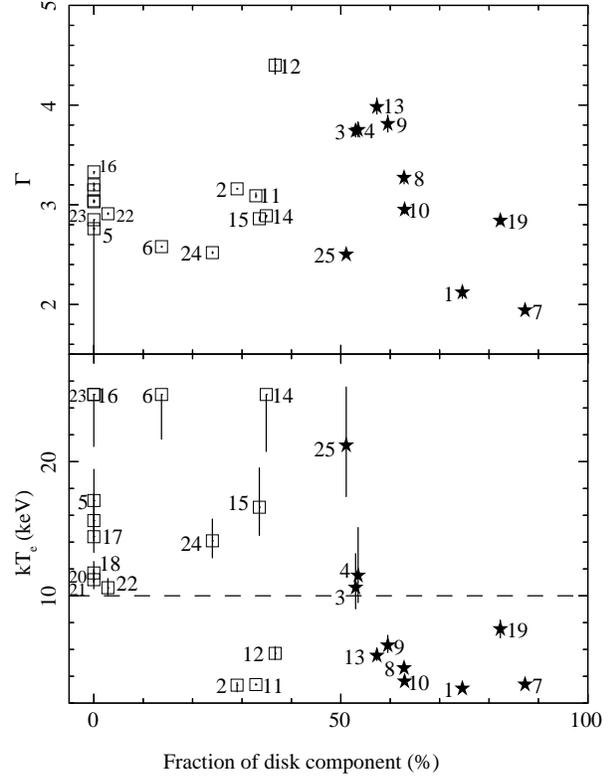,width=8.0cm,angle=0}}
  \end{center}
  \caption{$kT_{\rm e}$ and $\Gamma$ obtained from the fitting (see text). The dotted line in the lower panel represents $kT_{\rm e}=10$ keV.}\label{Figure:4}
\end{figure}

\subsection{Thermal Comptonization Component}
\begin{figure}
  \begin{center}
\centerline{\epsfig{file=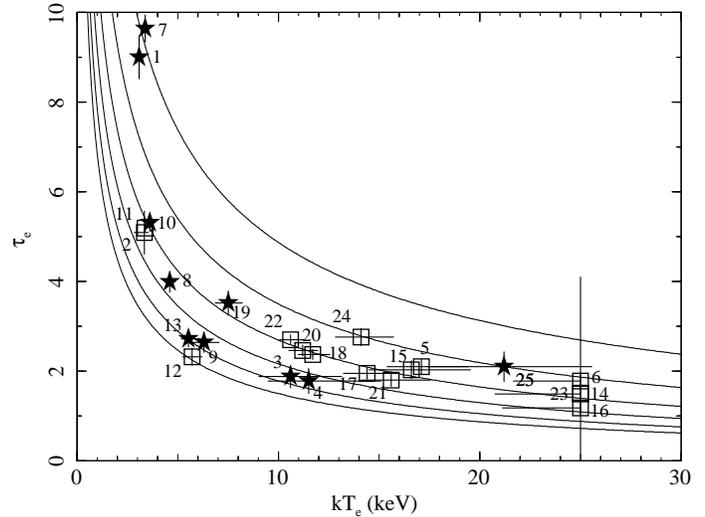,width=7.0cm,angle=270}}
  \end{center}
  \caption{$\tau$ vs. $kT_{\rm e}$ (see text). The lines are for constant value of $\Gamma$, that is $\Gamma=2,\ 2.5,\ 3,\ 3.5,\ 4,\ 4.5$, from top to bottom. The open squares are data with Comptonization-dominated spectra, while the filled stars are those with disk-dominated spectra.}\label{Figure:4}
\end{figure}
Here we will only focus on two important parameters from the
thermal Comptonization model that we use in our present study.
These parameters are the electron temperature, $kT_{\rm e}$, and the power-law
photon index, $\Gamma$.
The electron temperature determines the high energy rollover. In addition, the
low energy rollover is also considered and is parametrized by the seed photon
temperature. In our present study, this seed photon temperature is equal
to the inner disk temperature. The power-law photon index determines the slope
between low and high energy rollovers.

We will first describe the results for the case where the disk component is
modeled by MCD.
With some few exceptions, data with $>50\%$ disk fraction prefer low electron
temperature below 10 keV (see figure 8, filled stars).
We, however, set the maximum limit at 25 keV for the electron temperature
since, otherwise, it cannot be constrained by our data in some cases.
It is important to note here that the error estimate for the electron
temperature is, unlike other parameters, quite large.
In other words, we should be very careful in making an interpretation based
on the electron temperature results.
The power-law photon index varies between 2 -- 5 for most data regardless of
the disk fraction. Despite its relatively small errors, no clear trend is found
which left no room for further speculation (see figure 8).

In figure 9, the electron scattering optical depth, $\tau_{\rm e}$ is
calculated from $T_{\rm e}$ and $\Gamma$ by following Sunyaev and Titarchuk
(1980):
\begin{equation}
\tau_{\rm e}=\sqrt{2.25+\frac{3}{(T_{\rm e}/511 {\rm keV})[(\Gamma+0.5)^{2}-2.25]}}-1.5.
\end{equation}
The lines show constant $\Gamma=2$, 2.5, 3, 3.5, 4, and 4.5, from top to
bottom.
All data prefer an electron scattering optical depth between 1 and 10, mostly
greater than 2.

In the case where the disk component is modeled by extended DBB, the
electron temperatures tend to be higher but mostly still less than 10 keV,
in agreement with those of the MCD fits (see tables 3 and 4). 
The power-law photon index, $\Gamma$, remains within the same range as those
of the MCD fits although the values are not always similar.
We conclude that both the MCD and the extended DBB model prefer low
temperatures ($< 20$ keV) and thus optically thick ($\tau > 2$) corona
(since $\Gamma$ values do not vary so much).

\subsection{X-ray Hertzsprung-Russell (HR) Diagram}
\begin{figure}
  \begin{center}
\centerline{\epsfig{file=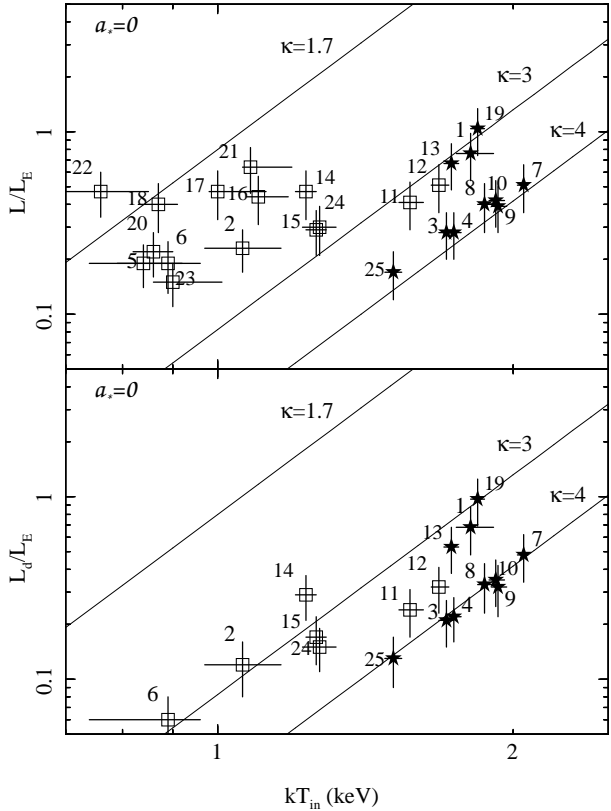,width=8cm,angle=0}}
  \end{center}
  \caption{Luminosity vs. disk temperature diagrams. We assumed $d=12.5$ kpc and $i=70^{\circ}$, and $M=14M_{\odot}$. The lines in each panel represent $\kappa=1.7$, $3.0$, and $4.0$ from top to bottom, respectively, for a non-rotating black hole (see text). Open squares are those with disk fraction $< 50 \%$ and filled stars for $>50\%$.} \label{Figure:10}
\end{figure}

One of the most general properties of accreting black hole systems is
that they follow a $L \propto T_{\rm eff}^{4}$ -track
(see, e.g. Gierli\'{n}ski and Done 2004) in a luminosity versus inner
disk temperature diagram (an X-ray Hertzsprung-Russell (HR) diagram) during
the high/soft state (i.e. when the disk component dominates the soft 
X-ray spectrum).
This property means that as the disk temperature and luminosity change, the
inner radius of the disk remains constant (Makishima et al. 1986; Ebisawa et
al. 1991; Dotani et al. 1997; Kubota et al. 2001).
Thus, the inner disk radius is often related to the innermost stable circular
orbit (ISCO) which in turn can be related to the black hole mass.

However, as the luminosity of the system approaches its Eddington luminosity,
it starts to depart from the $L \propto T_{\rm eff}^{4}$ -track.
Some systems, indeed, have shown that they start to follow a new flatter
track of $L \propto T_{\rm eff}^{2}$ (Kubota and Makishima 2004), which
satisfies the slim disk prediction for a non-rotating black hole.

Figure 10 shows the X-ray HR diagram of our sample.
Here we use $T_{\rm in}$ instead of $T_{\rm eff}$, but as long as we
consider a constant spectral hardening factor, $\kappa$,
$T_{\rm in} \propto T_{\rm eff}$.
The $y$-axis of the upper panel shows the total luminosity ($L$), obtained by
summing the luminosity of the disk component ($L_{\rm disk}$) and thermal
Comptonization component ($L_{\rm THC}$), normalized by the Eddington
luminosity.
The lower panel shows the disk luminosity normalized by the
Eddington luminosity.
The disk luminosity is calculated from the unabsorbed flux of MCD component
between 0.01 -- 100 keV assuming a disk geometry, a distance of 12.5 kpc, and
an inclination angle of 70$^{\circ}$.
We assume an isotropic emission when we calculate the thermal
Comptonization luminosity (also shown in table 3).
The filled stars show data with $>50\%$ disk fraction in their spectra,
while open squares represent the data with lower disk fraction ($<50\%$).
Data with $L_{\rm disk} < 0.02 L_{\rm E}$ are not shown in the lower panel
for clarity.

It is not always clear which
luminosity ($L$ or $L_{\rm disk}$) is better for this kind of plot. In the
present study we plot both luminosities for completeness.
In disk-dominated data, the disk fraction is moderately higher
than that of the thermal Comptonization component and thus both luminosities
give similar results. By contrast, for Comptonization-dominated data, their
position in the X-ray HR diagram depends on the choice of luminosity,
$L$ or $L_{\rm disk}$ (compare upper and lower panels of figure 10). 
As shown in the upper panel, these data have comparable total luminosities to
those of disk-dominated data. However, their disk fraction is mostly low
and thus their position is shifted to a lower position in the lower panel of
figure 10. Data with a very low disk fraction ($\sim 0$) do not follow
the $L_{\rm d} \propto T_{\rm in}^{4}$-track in the lower panel. Moreover, the
disk temperature value obtained from fitting becomes less reliable since the
disk contribution is almost negligible (see section 4.1.1).

From the lower panel of figure 10, we can see that there are two groups of
data following two different $L_{\rm d} \propto T_{\rm in}^{4}$-tracks.
The upper branch seems to extend to the lower disk luminosity data
(open squares) but these data are those whose spectra are dominated by
the thermal Comptonization.
All spectra of the data in the lower branch are
disk-dominated and their disk luminosity is below 0.5$L_{\rm E}$.
The lower branch may correspond to the highest end of standard accretion
disk branch in the high/soft state or slim-disk state (see below).

In figure 10, we also plot some lines from equation [9] of Makishima et al.
(2000), assuming $L_{\rm E}=1.5 \times 10^{38} (M/M_{\odot})$ erg s$^{-1}$.
It is assumed that the inner edge of the disk reaches the ISCO,
$r_{\rm in}=3r_{\rm S}$ for a
non-rotating black hole and $r_{\rm in}=0.5r_{\rm S}$ for a maximally
rotating black hole.
Given that we know the black hole mass from dynamical methods,
$M=14 M_{\odot}$, 
we can plot the lines for different values of $\kappa$, by assuming a value of
black hole spin.
In figure 10, we assume a black hole spin of 0 (non-rotating black hole).
These lines of constant $\kappa$ coincide with $L \propto T_{\rm in}^{4}$
-tracks.
The upper and lower branches, in the lower panel of figure 10, coincide with
$\kappa \sim 3$ and 4 for a non-rotating black hole, respectively.

\begin{figure}
  \begin{center}
\centerline{\epsfig{file=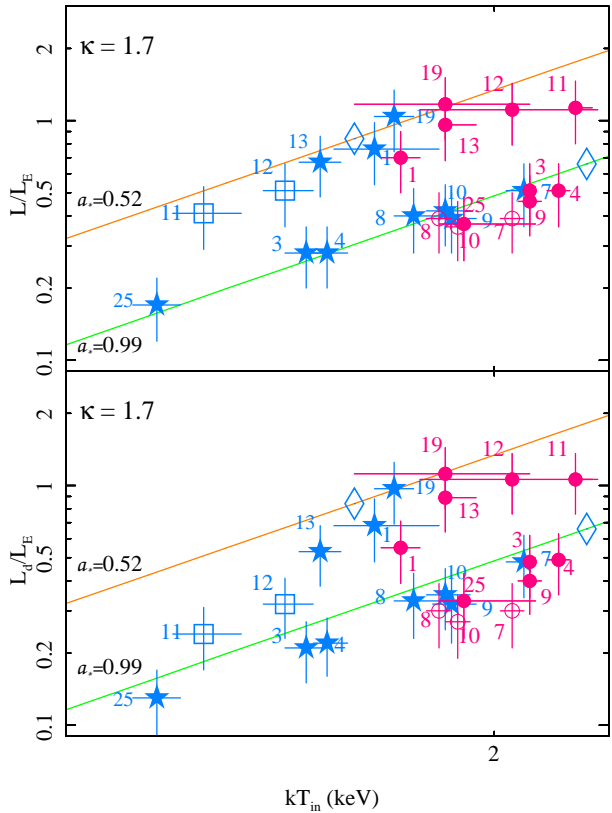,width=8cm,angle=0}}
  \end{center}
  \caption{X-ray HR diagrams for both MCD and extended DBB fits. Blue colors are from MCD fits: open squares for Comptonization-dominated data and filled stars for disk-dominated data. Red colors are from extended DBB fits: $p \sim 0.75$ shown as open circles, while $p<0.7$ as filled circles. Open diamonds show data from full-relativistic slim disk simulations where $\kappa$ is fixed at 1.7. The lines are $L \propto T_{\rm in}^{4}$-lines which cross each simulation data point (see text).} \label{Figure:5}
\end{figure}

Figure 11 shows the X-ray HR diagram for both MCD (blue points) and
extended DBB fits (red points).
The blue open squares and filled stars are Comptonization-dominated and
disk-dominated data, respectively.
Red filled circles are those datasets with $p<0.7$ from extended DBB fits,
while red open circles are those with $p \sim 0.75$.
We also plot the fitting results of two datasets from
numerical simulations of the slim disk model around a rotating black hole
(Sadowski 2009) in blue open diamonds. The higher temperature data
corresponds to a black hole spin of 0.99 while the lower temperature
to a black hole spin of 0.52. Relativistic effects
are also included when calculating the spectrum and the
spectral hardening factor is 1.7 (Sadowski et al. 2009). We draw
lines that cross each numerical simulation point (blue open diamond)
assuming $L \propto T_{\rm in}^{4}$ in orange and green.

It turns out that the dependence of $L$ on $T_{\rm in}$ becomes
less unclear because of the limited range of $T_{\rm in}$. Caution needs
to be taken, however, since the luminosity calculation from the
observed flux for the extended DBB model is less reliable than that of
the MCD model because the disk state is expected to become
the standard one and hence $p$ should change to 0.75 at outer radii.
Rather, we can directly compare our MCD fit results (blue points) with 
the same plot obtained from other BHBs (e.g. Figure~5 of Kubota
and Makishima 2004), where the $L \propto T_{\rm in}^{2}$-track 
is suggested at the highest temperature end based on the MCD model.

\section{Discussion}
GRS 1915+105 is known to be the only source in our Galaxy that persistently
shines at above 10\% of its Eddington luminosity.
It raises a doubt on whether we ever observed the canonical high/soft state 
even when the disk fraction is large.
We also found a puzzling behavior from our present study.
In this section we will discuss our interpretation as well as explore the
available scenarios for this puzzling behavior at near-Eddington luminosity.

\subsection{Accretion Disk States}
As shown in figure 7, there is a clear dependence of the MCD fitting
components, i.e. $r_{\rm in}$ and $kT_{\rm in}$ on the disk fraction.
As the disk fraction increases from 0 to 50\%, the innermost 
temperature of the disk increases as the radius decreases. 
It is also shown that $r_{\rm in}$ remains almost constant for those
datasets with disk fraction larger than 50\%. 

We suggest that the Comptonization-dominated spectra most probably
correspond to the (strong) very high state (VHS) found in other BHBs,
because of the similarity in: (1) the parameters
of Comptonization ($kT_{\rm e} \sim$10--20 keV and $\tau_{\rm e} \sim$
2--3), (2) the high $L/L_{\rm E}$ value, (3) the (apparently) large
$r_{\rm in}$ (or high $L$) for the given $T_{\rm in}$, and (4) the 
presence of strong (low frequency) QPOs (e.g. Morgan et al. 1997; Rodriguez
et al. 2004). These are
exactly found in the strong VHS of XTE~J1550--564, although the real
innermost radius may be smaller if the disk-corona coupling is taken
into account (Done and Kubota 2006). This VHS occurs when the mass
accretion rate becomes even higher than that of the high/soft
state. Theoretically, some outflow material blown off the disk due to
the increase of radiation pressure caused by the increase of mass
accretion rate is also predicted in this state.
It is very unlikely that the Comptonization-dominated spectra
correspond to the canonical low/hard state of BHBs, which is normally
observed at luminosities much lower than 10\% of the Eddington
luminosity, unlike our case. In addition, the electron temperature
of the Comptonizing corona in the low/hard state (e.g. Makishima et al.
2008) is much higher ($\sim$100 keV) than those found here.

We propose that the following three branches (states) have been 
observed in our data sample:
\begin{enumerate}
\item Comptonization-dominated spectra (when the disk flux
$< 30\%$ of the total flux), which correspond to the strong VHS 
in other BHBs.
\label {1}
\item Disk-dominated spectra: 
  \begin{enumerate}
  \item Lower luminosity branch: the high/soft state, or probably the slim disk
state, around a rotating black hole 
(see next subsection). 
  \label{a}
  \item Higher luminosity branch: probably a truncated disk with low
temperature ($<10$ keV) and optically thick Comptonization.
This would correspond to the new state that appears at luminosity very close
to the Eddington luminosity. It may be understood as the extension of the
high electron temperature VHS (state 1 above) when luminosity becomes higher.
  \label{b}
  \end{enumerate}
\label{2}
\end{enumerate}
State (1) and (2a) have already been observed in other BHBs, but state (2b)
looks unique in GRS 1915+105, probably due to its high $L/L_{\rm E}$. 
The disk-dominated spectra (state 2) will be discussed in more detail in
the following subsection. The cartoon of these three branches in the X-ray
HR diagram is shown in figure 12, which corresponds to the upper panel of
figure 10.
\begin{figure}
  \begin{center}
\centerline{\epsfig{file=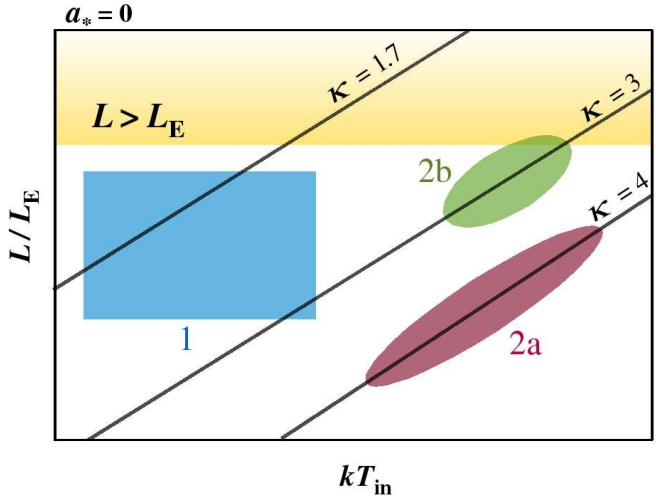,width=10.cm,angle=0}}
  \end{center}
  \caption{Cartoon of the three branches from the upper panel of figure 10.} \label{Figure:11}
\end{figure}

\subsection{On Disk-dominated Spectra and the Black Hole Spin}
The fact that the $L \propto T_{\rm in}^{2}$-track does not appear 
in the disk-dominated branches
even when the luminosity approaches and even exceeds the Eddington limit
contradicts the theoretical predictions of slim disk model for a non-rotating
black hole.
Moreover, as shown in the lower panel of figure 10, a high value of 
the spectral hardening factor, $\kappa \sim 4$ is required by the data in
the lower branch assuming a non-rotating black hole.
The higher value of $\kappa$ compared to the standard value commonly used for
the standard case $\kappa=1.7$ may be related to the increase in the mass
accretion rate.

Beloborodov (1998) has calculated the disk structure of a rapidly rotating
black hole with a super-Eddington accretion rate. The calculations show that
when the viscosity is high and the mass accretion rate approaches the critical
value, a strong deviation from equilibrium occurs in the inner region of the
disk. In this region, the plasma density is low and thus emission capability
is also low. The produced radiation in the disk is thus not fully thermalized
which results in higher disk temperature which further can be related to higher
value of $\kappa$. 
The question remains, however, since the luminosity of the lower branch, whose
$\kappa$ is higher, is lower than that of the upper branch.

We may have underestimated the luminosity of the lower branch.
For example, if some part of the disk is obscured by highly ionized outflow
materials which scatter disk radiation.
In the case when the outflow has a small opening angle, we should have
overestimated the luminosity due to high inclination angle of GRS 1915+105
and not the other way around.
If, however, we have a moderately geometrically thick disk (slim disk) together
with an outflow, there may be some interaction between the disk and the
outflow and complicate the radiation track. 
The material outflow from the disk is a delicate process which far from being
understood. 
We need a numerical calculation of disk-corona system with an outflow for
further analysis.
Kawashima et al. (2009) performed two dimensional radiation hydrodynamic
simulations of supercritical accretion flows with outflows.
They found a possible new state above the slim disk state in which both
disk and outflow components become important.
The slim disk and outflow model seems to be applicable in a particular
case of ULXs (Gladstone et al. 2009).

High inner disk temperatures found in our study may be explained by
a slim disk around a non-rotating black hole (e.g. Watarai et al. 2000),
a standard disk around a rotating black hole (e.g. Zhang et al. 1997),
or a slim disk around a rotating black hole (e.g. Sadowski 2009).
As mentioned above, the non existence of $L \propto T_{\rm in}^{2}$-track
and the too high $\kappa$ value contradict the theoretical prediction of
the slim disk around a non-rotating black hole.
While the rotating black hole is a good explanation for the
high inner disk temperatures (since the inner disk radius becomes closer
to the black hole), 
the luminosities of our sample are too high to keep the disk in the standard
accretion mode.

On the other hand, a clear deviation from the standard accretion disk
spectra during the high/soft state as luminosity approaches the Eddington
luminosity is in a good agreement with slim disk prediction.
Slim disk around a rotating black hole is one plausible
idea to resolve the above issues since the $L \propto T_{\rm in}^{2}$
-track is not necessarily held for the following reason.

In the case of a disk around a non-rotating black hole,
there is a gap between the radius of the ISCO (i.e., $3r_{\rm S}$) and that of
the event horizon ($1r_{\rm S}$).
As the mass accretion rate increases,
the flow density inside the ISCO increases and eventually fills the gap,
emitting significant radiation.
Hence, substantial radiation will be emitted from there
at luminosities near the Eddington.
This results in an apparent decrease of the inner edge of
the disk and increase of the innermost disk temperature
with an increase of the mass accretion rate
(and luminosity), in the way that $r_{\rm in} \propto T_{\rm in}^{-1}$.
This leads to the relation,
$L \propto r_{\rm in}^{2} T_{\rm in}^{4} \propto T_{\rm in}^{2}$
(Watarai et al. 2000).
This cannot happen for a maximally rotating black hole,
since the ISCO is located near the event horizon,
regardless of the mass accretion rate.
Hence, the relation, $L \propto T_{\rm in}^{2}$,
will never appear.

Furthermore, If we assume a rapidly rotating black hole, we do not need
to adopt too high spectral hardening factor for the data in the lower branch
of figure 11 (lower panel).
Provided that the spin parameter does not change, the nature
of the upper branch remains unclear .
One possible explanation is the change
of spectral hardening factor despite the unknown mechanism.
Another possible explanation is the effect of the strong Comptonization.
The inner disk is truncated and turns into Comptonizing corona,
which is suggested as a picture of the VHS of normal BHBs, and thus we observed
lower inner disk temperature than expected from high/soft state at the same
luminosity (Done and Kubota 2006).

The fact that the upper branch in our study has never been observed in other
Galactic microquasar, such as XTE J1550--564 and GRO J1655--40 (e.g. Kubota et
al. 2001; 2004), may have something to do with the spin parameter.
Both XTE J1550--564 and GRO J1655--40 are suggested to have a moderate spin
parameter $< 0.8$ (e.g. Gierli\'{n}ski et al. 2001; Davis et al. 2006;
Shafee et al. 2006).
Caution needs to be taken since there are several methods in
determining spin parameter which can lead to contradicting results.
In the case of GRO J1655-40, for example, Reis et al. (2009) obtained a lower
limit for the spin of 0.9 by using the reflection component in the spectrum.
At present, there is no consensus on the best methods to determine the black
hole spin parameter.
Despite the fact that our present study does not focus on
the determination of the spin parameter, our results suggest a
rapidly rotating black hole for GRS 1915+105.

McClintock et al. (2006) find a spin parameter $>0.98$ for GRS 1915+105
from their analysis.
They also find that the spin decreases as the luminosity increases. 
They, however, choose only data with $L_{\rm d}<30\%L_{\rm E}$
to determine the spin parameter.
The spin parameters obtained from data with $L_{\rm d}>40\%L_{\rm E}$, are
indeed in agreement with the moderate spin parameter obtained by Middleton
et al.(2006), spin parameter $\sim 0.7$. Middleton et al. (2006), however,
due to their selection criteria, have only a restricted data sample which
contains only high luminosity data, $L_{\rm d}>40\%L_{\rm E}$.
Comparing the disk luminosities and temperatures obtained from both Middleton
et al.(2006) and McClintock et al. (2006), data of Middleton et al. (2006)
seem to correspond to our higher branch, while those of McClintock et al.
(2006)
with $L_{\rm d}<30\%L_{\rm E}$ correspond to the lower branch in our X-ray
HR diagram.

\section{Conclusions}

We analyzed the spectra of GRS 1915+105 during its quasi-steady states
to study the supercritical accretion processes.
We found new interesting features as follows:
\begin{enumerate}
\item There is a strong dependence of the inner disk temperature and
radius on the fraction of the disk component: the disk temperature increases
as the disk fraction increases, while the opposite trend is found for the
inner disk radius. The electron temperature tends to be higher when the disk
fraction is small while the spectral slope remains more or less constant. 
\label{1}
\item We also fit the data with the extended DBB model to see any deviation from
the standard spectral shape. 
Despite the fact that the $p$-value deviates from
the standard value $p=0.75$ for data with $L>0.3L_{\rm E}$, the
$L \propto T_{\rm in}^{2}$-track predicted by the slim disk theory for a
non-rotating black hole does not appear.
A rapid spin of the black hole in GRS 1915+105 
can explain these features as well as a high value of spectral hardening
factor ($\sim 4$) that would be required for a non-rotating black hole.
\label{2}
\item We observed three spectral branches
(states) in our data sample: one Comptonization-dominated branch and two disk-
dominated branches. We suggest that Comptonization-dominated spectra
correspond to the canonical VHS observed in normal BHBs.
The lower luminosity branch of disk-dominated spectra seems to correspond
to the highest end of the high/soft state or the slim disk state around a
rotating black hole. The higher luminosity branch of disk-dominated
spectra may be interpreted as a truncated disk with low temperature and
optically thick Comptonization, unique to GRS 1915+105 at near Eddington
luminosity. 
\label{3}
\end{enumerate}

\bigskip
\bigskip
We thank the referee for their useful comments that helped improve this
work.
We gratefully thank Aleksander Sadowski, for providing the spectrum data from
numerical calculation of fully relativistic slim disk model, and Chris Done for
valuable discussion.
One of the authors, K.V., thank the Ministry of Education,
Culture, Sports, Science and Technology (MEXT) of Japan for the scholarship.
This work is supported in part by the Grant-in-Aid of MEXT (19340044, SM; 20540230, YU) and
by the Grant-in-Aid for the global COE programs on "The Next Generation of
Physics, Spun from Diversity and Emergence" from MEXT.


\begin{thebibliography}{62}
\bibitem[Belloni et al.(1997a)]{4}
   Belloni, T., M\'{e}ndez, M., King, A. R., van der Klis, M., \& van Paradijs, J. \ 1997a, \apj, 479, L145
\bibitem[Belloni et al.(1997b)]{5}
   Belloni, T., M\'{e}ndez, M., King, A. R., van der Klis, M., \& van Paradijs, J. \ 1997b, \apj, 488, L109
\bibitem[Belloni et al.(2000)]{6}
   Belloni, T., Klein-Wolt, M., M\'{e}ndez, M., van der Klis, M., \& van Paradijs, J. \ 2000, \aap, 355, 271
\bibitem[Beloborodov (1998)]{93}
   Beloborodov, A. M. \ 1998, \mnras, 297, 739
\bibitem[Castro-Tirado et al.(1992)]{8}
   Castro-Tirado, A. J., Brandt, S., \& Lund, N. \ 1992, \iaucirc, 5590, 2   
\bibitem[Castro-Tirado et al.(1994)]{9}
   Castro-Tirado, A. J., Brandt, S., Lund, N., Lapshov, I., Sunyaev, R. A., Shlyapnikov, A. A., Guziy, S., \& Pavlenko, E. P. \ 1994, \apjs, 92, 469 
\bibitem[Castro-Tirado et al.(1996)]{10}
   Castro-Tirado, A. J., Geballe, T. R., \& Lund, N. \ 1996, \apj, 461, L99
\bibitem[Chaty et al.(1996)]{7}
   Chaty, S., Mirabel, I. F., Duc, P. A., Wink, J. E., \& Rodr\'{i}guez, L. F. \ 1996, \aap, 310, 825
\bibitem[Davis et al.(2006)]{100}
   Davis, S. W., Done, C., \& Blaes, O. M. \ 2006, \apj, 647, 525
\bibitem[Dhawan et al.(2000)]{11}
   Dhawan, V., Goss, W. M., \& Rodr\'{i}guez, L. F. \ 2000, \apj, 540, 863
\bibitem[Done and Gierli\'{n}ski(2003)]{22}
   Done, C., \& Gierli\'{n}ski, M. \ 2003, \mnras, 342, 1041
\bibitem[Done et al.(2004)]{20}
   Done, C., Wardzi\'{n}ski, G., \& Gierli\'{n}ski, M. \ 2004, \mnras, 349, 393
\bibitem[Done and Kubota (2006)]{98}
   Done, C., \& Kubota, A. \ 2006, \mnras, 371, 1216
\bibitem[Dotani et al.(1997)]{88}
   Dotani, T., et al. \ 1997, \apjl, 485, 87
\bibitem[Ebisawa et al.(1991)]{87}
   Ebisawa, K., Mitsuda, K., \& Hanawa, T. \ 1991, \apj, 367, 213
\bibitem[Gierli\'{n}ski et al.(2001)]{99}
   Gierli\'{n}ski, M., Maciolek-Nied\'{z}wiecki, A., \& Ebisawa, K. \ 2001, \mnras, 325, 1253
\bibitem[Gierli\'{n}ski and Done (2004)]{86}
   Gierli\'{n}ski, M., \& Done, C. \ 2004, \mnras, 347, 885
\bibitem[Gladstone et al.(2009)]{95}
   Gladstone, J. C., Roberts, T. P., \& Done, C. \ 2009, \mnras, 397, 1836
\bibitem[Greiner et al.(2001a)]{12}
   Greiner, J., Cuby, J. G., McCaughrean, M. J., Castro-Tirado, A. J., \& Mennickent, R. E. \ 2001a, \aap, 373, L37
\bibitem[Greiner et al.(2001b)]{13}
   Greiner, J., Cuby, J. G., \& McCaughrean, M. J. \ 2001b, \nat, 414, 522
\bibitem[Hannikainen et al.(2005)]{104}
   Hannikainen, D. C., et al. \ 2005, \aap, 435, 995
\bibitem[Kaspi et al.(2000)]{73}
   Kaspi, S., Smith, P. S., Netzer, H., Maoz, D., Jannuzi, B. T., \& Giveon, U. \ 2000, \apj, 533, 631
\bibitem[Kato et al.(2008)]{71}
   Kato, S., Fukue, J., Mineshige, S. \ 2008, Black-Hole Accretion Disks (Kyoto: Kyoto University Press)
\bibitem[Kawashima et al.(2009)]{94}
   Kawashima, T., Ohsuga, K., Mineshige, S., Heinzeller, D., Takabe, H., \& Matsumoto, R. \ 2009, \pasj, 61, 769
\bibitem[Klein-Wolt et al.(2002)]{103}
   Klein-Wolt, M., Fender, R. P., Pooley, G. G., Belloni, T., Migliari, S., Morgan, E. H., \& van der Klis, M. \ 2002, \mnras, 331, 745
\bibitem[Kotani et al.(2000)]{80}
   Kotani, T., Ebisawa, K., Dotani, T., Inoue, H., Nagase, F., Tanaka, Y., \& Ueda, Y. \ 2000, \apj, 539, 413
\bibitem[Kubota et al.(1998)]{83}
   Kubota, A., Tanaka, Y., Makishima, K., Ueda, Y., Dotani, T., Inoue, H., \& Yamaoka, K. \ 1998, \pasj, 50, 667
\bibitem[Kubota et al.(2001)]{89}
   Kubota, A., Makishima, K., \& Ebisawa, K. \ 2001, \apj, 560, 147
\bibitem[Kubota and Makishima (2004)]{84}
   Kubota, A., \& Makishima, K. \ 2004, \apj, 601, 428
\bibitem[Magdziarz and Zdziarski (1995)]{52}
   Magdziarz, P., \& Zdziarski, A. A. \ 1995, \mnras, 273, 837
\bibitem [Makishima et al.(1986)]{42}
   Makishima, K., Maejima, Y., Mitsuda, K., Bradt, H. V., Remillard, R. A., Tuohy, I. R., Hoshi, R., \& Nakagawa, M. \ 1986, ApJ, 308, 635
\bibitem[Makishima et al.(2000)]{90}
   Makishima, K., et al. \ 2000, \apj, 535, 632
\bibitem[Makishima et al.(2008)]{102}
   Makishima, K., et al. \ 2008, \pasj, 60, 585
\bibitem[McClintock et al.(2006)]{81}
   McClintock, J. E., Shafee, R., Narayan, R., Remillard R. A., Davis, S. W., \& Li, L. \ 2006, \apj, 652, 518
\bibitem[Middleton et al.(2006)]{21}
   Middleton, M., Done, C., Gierli\'{n}ski, M., \& Davis, S. W. \ 2006, \mnras, 373, 1004
\bibitem[Mineshige et al.(1994)]{44}
   Mineshige, S., Hirano, A., Kitamoto, S., Yamada, T. T., \& Fukue, J. \ 1994, \apj, 426, 308
\bibitem[Mirabel and Rodr\'{i}guez(1994)]{19}
   Mirabel, I. F., \& Rodr\'{i}guez, L. F. \ 1994, \nat, 371, 46
\bibitem[Mirabel et al.(1998)]{24}
   Mirabel, I. F., Dhawan, V., Chaty, S., Rodr\'{i}guez, L. F., Marti, J., Robinson, C. R., Swank, J., \& Geballe, T. \ 1998, \aap, 330, L9
\bibitem[Mitsuda et al.(1984)]{41}
   Mitsuda, K., et al. \ 1984, \pasj, 36, 741
\bibitem[Morgan et al.(1997)]{16}
   Morgan, E. H., Remillard, R. A., \& Greiner, J. \ 1997, \apj, 482, 993
\bibitem[Neilsen and Lee(2009)]{23}
   Neilsen, J., \& Lee, J. C. \ 2009, \nat, 458, 481
\bibitem[Ohsuga et al.(2009)]{70}
   Ohsuga, K., Mineshige, S., Mori, M., \& Kato, Y. \ 2009, \pasj, 61, 7
\bibitem[Okajima et al.(2006)]{76}
   Okajima, T., Ebisawa, K., \& Kawaguchi, T. \ 2006, \apjl, 652, 105
\bibitem[Orosz et al.(2007)]{14}
   Orosz, J. A. \ 2007, \nat, 449, 872
\bibitem[Poutanen et al.(2007)]{78}
   Poutanen, J., Lipunova, G., Fabrika, S., Butkevich, A.G., \& Abolmasov, P. \ 2007, \mnras, 377, 1187	
\bibitem[Prestwich et al.(2007)]{15}
   Prestwich, A. H., Kilgard, R., Crowther, P. A., Carpano, S., Pollock, A. M. T., Zezas, A., Saar, S. H., Roberts, T. P., \& Ward, M. J. \ 2007, \apj, 669, L21
\bibitem[Reis et al.(2009)]{105}
   Reis, R.C., Fabian, A. C., Ross, R. R., \& Miller, J. M. \ 2009, \mnras, 395, 1257
\bibitem[Rodriguez et al.(2004)]{101}
   Rodriguez, J., Corbel, S., Hannikainen, D. C., Belloni, T., Paizis, A., \& Vilhu, O. \ 2004, \apj, 615, 416
\bibitem[Rodriguez et al.(2008a)]{17}
   Rodriguez, J., et al. \ 2008a, \apj, 675, 1436
\bibitem[Rodriguez et al.(2008b)]{18}
   Rodriguez, J., et al. \ 2008b, \apj, 675, 1449
\bibitem[Sadowski (2009)]{91}
   Sadowski, A. \ 2009, \apjs, 183, 171
\bibitem[Sadowski et al. (2009)]{92}
   Sadowski, A., Abramowicz, M. A., Bursa, M., Klu\'{z}niak, W., R\'{o}$\dot{\rm z}$a\'{n}ska, A., \& Straub, O. \ 2009, \aap, 502, 7
\bibitem[Shafee et al.(2006)]{97}
   Shafee, R., McClintock, J. E., Narayan, R., Davis, S. W., Li, L. \& Remillard, R. A. \ 2006, \apj, 636, L113
\bibitem[Shakura and Sunyaev(1973)]{43}
   Shakura, N. I., \& Sunyaev, R. A. \ 1973, \aap, 24, 337
\bibitem[Sunyaev and Titarchuk (1980)]{85}
   Sunyaev, R. A., \& Titarchuk, L. G. \ 1980, \aap, 86, 121
\bibitem[Takeuchi et al.(2009)]{79}
   Takeuchi, S., Mineshige, S., \& Ohsuga, K. \ 2009, \pasj, 61, 783
\bibitem[Tsunoda et al.(2006)]{77}
   Tsunoda, N., Kubota, A., Namiki, M., Sugiho, M., Kawabata, K., \& Makishima, K. \ 2006, \pasj, 58, 1081
\bibitem[Ueda et al.(2009)]{74}
   Ueda, Y., Yamaoka, K., \& Remillard, R. \ 2009, \apj, 695, 888
\bibitem[Vierdayanti et al.(2006)]{75}
   Vierdayanti, K., Mineshige, S., Ebisawa, K. \& Kawaguchi, T. \ 2006, \pasj, 58, 915
\bibitem[Wandel et al.(1999)]{72}
   Wandel, A., Peterson, B. M., \& Malkan, M. A. \ 1999, \apj, 526, 579
\bibitem[Watarai et al.(2000)]{96}
   Watarai, K., Fukue, J., Takeuchi, M., \& Mineshige, S. \ 2000, \pasj, 52, 133
\bibitem[Zhang et al.(1997)]{82}
   Zhang, S. N., Cui, W., \& Chen, W. \ 1997, \apj, 482, L155
\bibitem[Zycki et al.(1999)]{45}
   Zycki, P. T., Done, C., \& Smith, D. A. \ 1999, \mnras, 309, 561
\end{thebibliography}
\end{document}